\documentclass[a4paper, 10pt]{article}
\pdfoutput=1

\usepackage[left=2cm,right=2cm,top=2cm,bottom=2cm]{geometry}
\usepackage{graphicx}
\usepackage{amsmath}
\usepackage[utf8]{inputenc}
\usepackage{hyperref}
\usepackage{url}
\usepackage{authblk}
\usepackage{amssymb}
\usepackage{amsfonts}
\usepackage{indentfirst}
\usepackage[toc,title]{appendix}
\usepackage[font=small]{caption}
\usepackage{color}

\newcommand{\tr}{\textup{Tr}}
\newcommand{\btr}{\textup{bTr}}
\newcommand{\<}{\left<}
\renewcommand{\>}{\right>}
\newcommand{\idm}{\mathbf{1}}
\newcommand{\zb}{\bar{z}}
\newcommand{\wb}{\bar{w}}

\newcommand{\cG}{{\cal G}}
\newcommand{\cQ}{{\cal Q}}
\newcommand{\cX}{{\cal X}}

\newcommand{\cH}{{\cal H}}
\newcommand{\cP}{{\cal P}}
\newcommand{\cK}{{\cal K}}

\newcommand{\ket}[1]{\left| #1 \>}
\newcommand{\bra}[1]{\< #1 \right|}
\newcommand{\braket}[2]{\< #1 | #2 \>}

\newcommand{\re}{\textup{Re}}
\newcommand{\im}{\textup{Im}}
\newcommand{\ob}{\bar{1}}
\newcommand{\diag}{\textup{diag}}

\begin{document}

\title{Probing non-orthogonality of eigenvectors in non-Hermitian matrix models: diagrammatic approach}
\author[1]{Maciej A. Nowak\thanks{maciej.a.nowak@uj.edu.pl} }
\author[1]{Wojciech Tarnowski\thanks{wojciech.tarnowski@uj.edu.pl}}
\affil[1]{M. Smoluchowski Institute of Physics and
Mark Kac Complex Systems Research Center, Jagiellonian University,
S. Łojasiewicza 11,
PL 30-348 Kraków, Poland.}
\date{\today}
\maketitle

\begin{abstract} Using large $N$  arguments,  we propose a scheme for calculating the two-point eigenvector correlation function for  non-normal random matrices in the large $N$ limit. The setting generalizes the quaternionic extension of free probability to two-point functions. In the particular case of biunitarily invariant random matrices, we obtain a simple, general expression for the two-point eigenvector correlation function, which can be viewed as a further  generalization of the single ring theorem. This construction has some striking similarities to the  freeness of the second kind known for the Hermitian ensembles in large $N$. On the basis of several solved examples, we conjecture two kinds of microscopic universality of the eigenvectors - one in the bulk,  and one at the rim.  The form of the conjectured bulk universality agrees  with the  scaling limit found by Chalker and Mehlig [JT Chalker, B Mehlig, PRL, \textbf{81}, 3367 (1998)] in the case of the complex Ginibre ensemble. 
\end{abstract}


\section{Introduction}

Non-normal operators are ubiquitous in physical models. Examples include hydrodynamics, open quantum systems, PT-symmetric Hamiltonians, Dirac operators  in the presence of a chemical potential  or finite angle $\theta$.  Non-normality is responsible for the transient dynamics, sensitivity of the spectrum to perturbations, pseudoresonant behavior and rapid growth of the perturbations of the system~\cite{trefethen2005pseudospectra}. These effects are relevant in plasma physics~\cite{PlasmaPhysics}, fluid mechanics~\cite{TransientFluidMechanics}, ecology~\cite{Resilience,asllani2017topological}, laser physics~\cite{Siegman1995}, atmospheric science~\cite{Atmospheric}, and magnetohydrodynamics~\cite{magnetohydro}, just to mention a few.
Non-normality is common in dynamical systems as its simplest source is the asymmetry of coupling between components.

Historically, most of the studied properties of  non-normal random operators  dealt with the eigenvalues. The eigenvalues of such operators are usually  complex, requiring new  calculational techniques, at the level of both macroscopic and microscopic correlations.  Surprisingly, this quest  for complex eigenvalues has eclipsed the study of eigenvectors, which are perhaps most  distinctive features of non-normal  operators. In particular, non-normal operators have two sets of eigenvectors, left and right,   which  are non-orthogonal among themselves,  but can be chosen to be  bi-orthogonal, provided  that  the non-normal operator  can be diagonalized at all.

One of the first attempts  to develop a systematic understanding of the non-orthogonality of eigenvectors in non-Hermitian random matrices was made by Chalker and Mehlig~\cite{chalker1998correlator,mehlig2000correlator}. Despite their study concentrated on the complex Ginibre ensemble, perhaps the simplest non-normal random operator,   the  results  turned out quite non-trivial and revealed the possibilities of well-hidden universal properties  of eigenvectors of non-normal operators. Another connection   of  the properties of non-normal operators  and their eigenvectors   to free probability was  established  soon after~\cite{correlator1999quaternionic},   but the systematic study  of  this topic  has not followed.   Only very recently,  the topic of eigenvectors of non-normal operators was picked back up. First, the transient growth driven by eigenvector non-orthogonality was proposed as a mechanism of amplification of neural signals in balanced neural networks~\cite{MurphyBalancedAmplification,VogelsNonnormal,HennequinTransient} and giant amplification of noise crucial in the formation of Turing patterns~\cite{AmplificationTuring,TransientTuringPatterns,KlikaTransientTuring}. Second, the non-orthogonality of eigenfunctions was related to the statistics of resonance width shifts in open quantum systems~\cite{SavinFyodorov}, which was soon confirmed experimentally~\cite{Legrand}.
Third, the  essential role of eigenvectors in stochastic motion of eigenvalues was revealed~\cite{Movassagh2016,DysonianDynamics,DiffusionExtended}.  Last but not least, the topic has triggered the attention of the mathematical community~\cite{WaltersStarr,BourgadeDubach}. 

In this work we focus on statistical ensembles of complex non-Hermitian matrix models, the probability density of which is invariant under the action of the unitary group $P(X)=P(UXU^{\dagger})$. We also assume that in the $N\to\infty$ limit, at which we are working, the eigenvalues of $X$ concentrate on a compact domain of a complex plane. Our results are valid for $|z-w|$ of order 1. We will study  one-point and two-point Green functions  built out of left and right eigenvectors.    
Here we recall, that if a non-normal matrix $X$ can be diagonalized by a similarity transformation $X=S\Lambda S^{-1}$, it possesses two eigenvectors for each eigenvalue $\lambda_i$: right $\ket{R_i}$ (a column in the matrix notation) and left $\bra{L_i}$ (row), satisfying the eigenequations
\begin{equation}
X\ket{R_i}=\lambda_i\ket{R_i}, \quad \bra{L_i}X=\bra{L_i}\lambda_i.
\end{equation}
These eigenvectors are not orthogonal $\braket{L_i}{L_j}\neq\delta_{ij}\neq\braket{R_i}{R_j}$, but normalized by the biorthogonality condition $\braket{L_i}{R_j}=\delta_{ij}$. They also satisfy the completeness relation $\sum_{k=1}^{N}\ket{R_k}\bra{L_k}=\idm$. These two properties leave a freedom of rescaling each eigenvector by a non-zero complex number, $\ket{R_i}\to c_i\ket{R_i}$ with $\bra{L_i}\to\bra{L_i}c_{i}^{-1}$. They also allow for multiplication by a unitary matrix $\ket{R_i}\to U\ket{R_i}$ and $\bra{L_i}\to\bra{L_i}U^{\dagger}$. Upon the second transformation the new vectors are not the eigenvectors of the original matrix but of  one given by the adjoint action of the unitary group $X\to UXU^{\dagger}$, which suggests that a natural probability measure should assign these two matrices the same probability density function (pdf). The simplest object, which is invariant under these transformations, is the matrix of overlaps $O_{ij}=\braket{L_i}{L_j}\braket{R_j}{R_i}$~\cite{chalker1998correlator,mehlig2000correlator}.

To see how the eigenvector correlation functions appear naturally, let us consider an average $\<\frac{1}{N}\tr f(X)g(X^{\dagger})\>$, where $f,g$ are two functions analytic in the spectrum of $X$ and $\<f(X)\>=\int f(X) P(X)dX$ denotes the average with respect to the pdf $P(X)$. Taking $f=g$, we get the (normalized) Frobenius norm of a function of  matrix. The $1/N$ normalization was taken to get a finite quantity in the $N\to\infty$ limit. Using the spectral decomposition $X=\sum_{k=1}^{N}\ket{R_k}\lambda_k\bra{L_k}$ and inserting the identity, $1=\int d\mu(z)\delta^{(2)}(z-\lambda_k)$ twice, we obtain the expression
\begin{equation}
\<\frac{1}{N}\tr f(X)g(X^{\dagger})\>=\int d\mu(z) d\mu(w) f(z) g(\wb) D(z,w), \label{eq:Average}
\end{equation}
with
\begin{equation}
D(z,w)=\<\frac{1}{N}\sum_{k,l=1}^{N} O_{kl}\delta^{(2)}(z-\lambda_k)\delta^{(2)}(w-\lambda_l)\>. \label{eq:Density}
\end{equation}
The two dimensional Dirac delta is understood as two deltas for real and imaginary parts $\delta^{(2)}(z)=\delta(\re z)\delta(\im z)$, and the measure $d\mu(z)=dx dy$ for $z=x+iy$. $D(z,w)$ introduced in~\cite{chalker1998correlator,mehlig2000correlator} is the density of eigenvalues weighted by the invariant overlap of the corresponding eigenvectors. It is split into a regular and singular part $D(z,w)=\tilde{O}_1(z)\delta^{(2)}(z-w)+O_{2}(z,w)$, where
\begin{equation}
\tilde{O}_1(z)=\<\frac{1}{N}\sum_{k=1}^{N}O_{ii}\delta^{(2)}(z-\lambda_i)\>,\qquad O_2(z,w)=\<\frac{1}{N}\sum_{\substack{k,l=1 \\ k\neq l}}^{N}O_{kl}\delta^{(2)}(z-\lambda_k)\delta^{(2)}(w-\lambda_l)\>. \label{eq:CorrFuncDef}
\end{equation}
A one-point function, defined this way, in the bulk and far from the rims of the complex spectra grows linearly with the size of a matrix. To have a finite limit in large $N$, one considers the scaled function $O_1(z)=\frac{1}{N}\tilde{O}_1(z)$. Throughout the paper we shall use only the 'untilded' function.

The one-point function $O_1$ plays an important role in scattering in open chaotic cavities~\cite{FrahmSchomerus,SavinFyodorov} and random lasing~\cite{Schomerus,Patra}, where the  so-called Petermann factor~\cite{Petermann} modifies the quantum-limited linewidth of a laser. It is also crucial in the description of the diffusion processes on matrices~\cite{DysonianDynamics,DiffusionExtended} and gives the expectation of the squared eigenvalue condition number~\cite{CondNum}, an important quantity governing the stability of eigenvalues~\cite{trefethen2005pseudospectra,wilkinson1965algebraic}. The exact calculations are possible for Gaussian matrices~\cite{chalker1998correlator,mehlig2000correlator,WaltersStarr}, in the matrix model for open chaotic scattering~\cite{Schomerus,Patra,FyodorovMehlig} and for products of small Gaussian matrices~\cite{BurdaVivo}. For the Ginibre matrix the full distribution of the diagonal overlap is available and turns out to be heavy-tailed, as discovered by Bourgade and Dubach~\cite{BourgadeDubach} with the use of probabilistic techniques, and investigated later using integrable structure and sypersymmetry by Fyodorov~\cite{FyodorovOverlap}.

Despite that the overlap between eigenvectors are crucial in the description of the dynamic of the linear system~\cite{MartiBrunelOstojic} and in the decay laws in open quantum systems~\cite{SavinSokolov}, the two-point function is much less known. The exact results are obtained only for the Ginibre matrix~\cite{chalker1998correlator,mehlig2000correlator,BourgadeDubach}  and for open chaotic scattering with a single channel coupling~\cite{FyodorovMehlig}. Even the asymptotic results are known only for Gaussian matrices~\cite{chalker1998correlator,mehlig2000correlator} and the quantum scattering ensemble~\cite{MehligSanterQuantumScattering}. The aim of this paper is to extend the known asymptotic results and develop a diagrammatic technique for calculation of the two-point function in the large $N$ limit. 

The paper is organized as follows. 
In Section~\ref{sec:Reminder} we briefly recall the cornerstones of diagrammatic calculus~\cite{correlator1999quaternionic} for one-point Green's functions in the large non-Hermitian ensembles, to show an analogy between the formalism developed in this paper and the diagrammatic approach to one-point functions. This encapsulates both the mean spectral density and the one-point eigenvector correlation function $O_1$.  Appendix~\ref{sec:Elliptic}  shows concrete calculations within this formalism for the elliptic ensemble. 

Section~\ref{Sec:Main} contains the main body of the paper -- a formalism for the calculation of $O_2$ in the large $N$ limit. We extend the diagrammatic technique introduced by Chalker and Mehlig for Gaussian matrices to any probability distribution with unitary symmetry. Regularizing and linearizing the product of resolvents, we embed them into the quaternionic space.  The analysis of planar Feynman diagrams leads us to the matrix Bethe-Salpeter equation \eqref{eq:Master1}, which relates the product of resolvents with the one-point Green's function and planar cumulants. The latter are encoded in their generating function -- quaternionic $R$-transform, see \eqref{eq:Apowerseries}.  As a result, the two-point eigenvector correlation function is completely determined by the one-point functions encoding the  spectral density and $O_1$. This result resembles the Ambj{\o}rn-Jurkiewicz-Makeenko universality for Hermitian ensembles~\cite{AmbjornJurkiewiczMakeenko}.

We also study the traced product of resolvents $\mathfrak{h}(z,\wb)=\<\frac{1}{N}\tr (z\idm-X)^{-1}(\wb\idm-X^{\dagger})^{-1}\>$, which allows for the calculation of the average \eqref{eq:Average} as a Dunford-Taylor integral~\cite{Dunford1957,HornJohson1990}
\begin{equation}
\<\frac{1}{N}\tr f(X)g(X^{\dagger})\>=\frac{1}{(2\pi i)^2}\int_{\gamma}dz\int_{\bar{\gamma}}d\wb f(z)g(\wb)\mathfrak{h}(z,\wb),
\end{equation}
where contours $\bar{\gamma}$,$\gamma$ encircle all eigenvalues of $X$ clockwise and counterclockwise, respectively. We derive the equation for $\mathfrak{h}$, expressing it in terms of quaternionic $R$-transform and traced resolvents, see~\eqref{eq:twopoint} and \eqref{eq:A11}.

An important and still quite large subclass of non-Hermitian ensembles for which the main equations \eqref{eq:Master1}\eqref{eq:Apowerseries} admit further simplifications consists of matrices, the pdf of which is invariant under the transformation by two independent unitary matrices $U,V\in U(N)$, i.e. $P(X)=P(UXV^{\dagger})$, thus called the biunitarily invariant ensemble~\cite{KieburgKosters}. In this case we obtain a compact formula for the two-point eigenvector correlation function 

\begin{equation}
O_2(z_1,z_2)=\frac{1}{\pi} \partial_{\zb_1}\partial_{z_2}\frac{\zb_1 (z_1-z_2)O_1(r_1)+z_2(\zb_1-\zb_2) O_1(r_2)}{|z_1-z_2|^2 \left[F(r_1)-F(r_2)\right]}. 
\end{equation}
Here $F$ is the radial cumulative distribution function (cdf), defined as $F(r)=2\pi\int_0^r\rho(s) sds$, with $\rho(s)$ the spectral density circularly symmetric on the complex plane. The one-point eigenvector function is related to $F$ via~\cite{CondNum}
\begin{equation}
O_1(r)=\frac{F(r)(1-F(r))}{\pi r^2}, \label{eq:O1Rdiag}
\end{equation}
and $r=|z|$. The traced product of resolvents is shown to take a universal form
\begin{equation}
\mathfrak{h}(z,\wb)=\frac{1}{z\wb-r_{out}^2},
\end{equation}
where $r_{out}$ is the spectral radius. This result has been already obtained for the Ginibre ensemble~\cite{mehlig2000correlator} and, recently, for matrices with independent identically distributed (iid) entries~\cite{erdos2017power}.

Applications of the developed formalism are presented in Sec.~\ref{sec:Examples}, where we consider an elliptic ensemble, some  instances from the biunitarily invariant class: truncated unitary, induced Ginibre, the product of two Ginibres and their ratio. As the last example we consider a pseudohermitian matrix -- a product of two shifted GUE matrices.
In Section~\ref{sec:Micro}, we discuss the consequences of our large $N$ results on the microscopic regime.  We conjecture, on the basis of the few examples solved in the literature and using our own results, that the two-point eigenvector correlation functions may exhibit universal bulk scaling, as what happens for the microscopic  spectral two-pointers in Hermitian matrix models. More precisely, we conjecture that in generic complex non-Hermitian matrices for all points in the bulk at which the spectral density does not develop singularities there exists a limit
\begin{equation}
\lim_{N\to\infty}N^{-2}O_2(z+\frac{x}{\sqrt{N}},z+\frac{y}{\sqrt{N}})=O_1(z) \Phi(|x-y|),
\end{equation}
where 
\begin{equation}
\Phi(|\omega|)=-\frac{1}{\pi^2|\omega|^4}\left(1-(1+|\omega|^2)e^{-|\omega|^2}\right).\label{eq:Micro}
\end{equation}
Section~\ref{sec:Conclusions} concludes the paper and points at some possible further developments.

\section{Non-Hermitian random matrices}
\label{sec:Reminder}
In non-Hermitian random matrix theory one is primarily interested in the distribution of the eigenvalues $\rho(z)=\<\frac{1}{N}\sum_{i=1}^{N}\delta^{(2)}(z-\lambda_i)\>$. The 2-dimensional Dirac delta can be recovered using the relation $\partial_{\zb}\frac{1}{z}=\pi \delta^{(2)}(z)$. Unfortunately, the average over the ensemble of the trace of the resolvent $\mathfrak{g}(z)=\<\frac{1}{N}\tr (z\idm-X)^{-1} \>$ does not yield the correct result inside the spectrum, as one would naively expect. The reason for this failure is that differentiation and taking the ensemble average are not interchangeable. This phenomenon was termed the  spontaneous breaking of holomorphic symmetry~\cite{nonhermitianmatrixmodels}.

A way to circumvent this obstacle is to consider a regularization of the Dirac delta. In RMT one mostly considers the 2D Poisson kernel
\begin{equation}
\pi\delta^{(2)}(z)=\lim_{\epsilon\to 0} \frac{\epsilon^2}{(|z|^2+\epsilon^2)^2}=\lim_{\epsilon\to 0} \partial_{\zb} \frac{\zb}{|z|^2+\epsilon^2}.
\end{equation} 
The expression on the right hand side provides a prescription for how the resolvent in the spectrum of $X$ should be regularized. Having this hint in mind, one defines
\begin{equation}
g(z,\zb,w,\wb)=\<\frac{1}{N}\tr (\zb\idm -X^{\dagger})[(z\idm-X)(\zb\idm -X^{\dagger})+|w|^2 \idm ]^{-1} \>. \label{eq:g}
\end{equation}
The spectral density can be now calculated via
\begin{equation}
\rho(z,\zb)=\frac{1}{\pi} \lim_{|w|\to 0} \partial_{\zb} g(z,\zb,w,\wb),  \label{eq:SpecDens}
\end{equation}
which can be also understood as a Poisson law in 2D electrostatics, since
\begin{equation}
\rho(z,\zb)=\lim_{|w|\to 0}\frac{1}{\pi} \partial_z\partial_{\zb} \Phi(z,\zb,w,\wb),
 \end{equation}
 where 
 \begin{equation}
\Phi(z,\zb,w,\wb)=\<\frac{1}{N}\ln\det [(z\idm-X)(\zb\idm-X^{\dagger})+|w|^2 \idm ]\> \label{eq:Potential}
 \end{equation}
is the (regularized) electrostatic potential of charges interacting via repulsive central force $F(r)\sim \frac{1}{r}$.

\subsection{Linearization}
Due to the quadratic expression in $X$ in the denominator, the average in \eqref{eq:g} is intractable when non-normal matrices are considered. To circumvent this problem one introduces the $2N\times 2N$ matrix~\cite{nonhermitianmatrixmodels,nonhFRV,feinbergZee,ChalkerWang}
\begin{equation}
\cG=\<\left(\begin{array}{cc}
z\idm -X & i\wb \idm \\
i w \idm & \zb\idm -X^{\dagger}
\end{array}\right)^{-1}\> \label{eq:QuatResolvent}
\end{equation}
and the block trace operation, mapping $2N\times 2N$ matrices onto $2\times 2$ ones
\begin{equation}
\btr \left(\begin{array}{cc}
A & B \\
C & D 
\end{array}\right)=\left(\begin{array}{cc}
\tr A & \tr B \\
\tr C & \tr D
\end{array} \right).
\end{equation}
Then, one defines the $2\times 2$ Green's function
\begin{equation}
G(z,\zb,w,\wb)=\left(\begin{array}{cc}
G_{11} & G_{1\ob} \\
G_{\ob 1} & G_{\ob\ob}
\end{array}\right)=\frac{1}{N}\btr \cG (z,\zb,w,\wb).
\end{equation}
Its upper-left entry is exactly the desired  function $g$ (cf. \eqref{eq:g}). Once Green's function is known, one gets four elements of $G$. The entry $G_{\ob\ob}$ is just the complex conjugate of $G_{11}$, thus does not provide any additional information. The off-diagonal entries $G_{1\ob}=-\bar{G}_{\ob 1}$ in the large $N$ limit give the one-point eigenvector correlation function~\cite{correlator1999quaternionic}
\begin{equation}
O_1(z)=\lim_{|w|\to 0}-\frac{1}{\pi}G_{1\ob}G_{\ob 1}. \label{eq:Correlator}
\end{equation}

\subsection{Quaternionic structure}
Green's function can be conveniently written as
\begin{equation}
G=\< \frac{1}{N}\btr (\cQ-\cX)^{-1} \>=\left(\begin{array}{cc}
\partial_{Q_{11}} \Phi & \partial_{Q_{\ob 1}}\Phi  \\
\partial_{Q_{1\ob}}\Phi & \partial_{Q_{\ob\ob}}\Phi
\end{array}\right). \label{eq:GreenQuaternion}
\end{equation}
with 
\begin{equation}
\cX=\left(\begin{array}{cc}
X & 0 \\
0 & X^{\dagger}
\end{array}\right), \quad
\cQ=Q\otimes \idm, \quad
Q=\left(\begin{array}{cc}
z & i\wb \\
i w & \zb 
\end{array}\right).
\end{equation}
This form of  Green's function resembles its traditional form as a traced resolvent, but now its argument is a $2\times 2$ matrix and the matrix $X$ is duplicated to incorporate also $X^{\dagger}$. The matrix $Q$ is a representation of a quaternion $q=x+iy+ju+kv$ with the identification $z=x+iy$ and $w=v+iu$~\cite{Cayley}. The entries of $G$ satisfy the same algebraic constraints as $Q$, therefore $G$ is itself a quaternion and we refer to it as the quaternionic Green's function, similarly $\cG$ is called the quaternionic resolvent. 

\subsection{Averages in large N}
\begin{figure}\begin{center}
\includegraphics[width=0.4\textwidth]{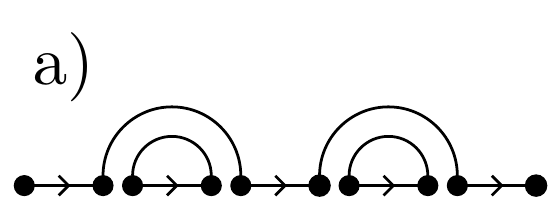} \includegraphics[width=0.4\textwidth]{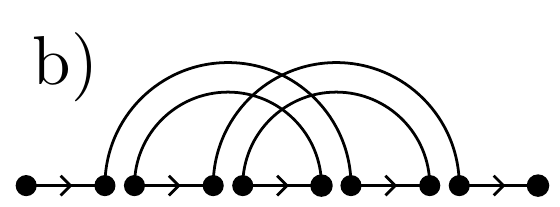}
\end{center}
\caption{ Examples of planar (a) and non-planar (b) diagrams in the diagrammatic expansion of the Gaussian model coming from the term $\<\cQ^{-1}(\cX\cQ^{-1})^4\>$. For a general matrix model with an arbitrary potential the order of the diagram is given by $N^{L+V-P}$, where $V$ is the number of vertices, $L$ is the number of loops and $P$ is the number of propagators comprising the diagram. This shows that the dominant contribution comes from the planar diagrams. The contribution from the non-planar diagram on (b) is of order $N^{-2}$, thus vanishes in the large $N$ limit. 
\label{Fig:PlanarDiagrams}
}
\end{figure}

We are interested in calculations of the averages of some functions of $X$, e.g. $\<f(X,X^{\dagger})\>$, with respect to distributions invariant under the adjoint action of the unitary group $P(X)=P(UXU^{\dagger})$. We parameterize them by
\begin{equation}
P(X)\sim \exp\left(-N\tr V(X,X^{\dagger})\right).
\end{equation}
$V(X,X^{\dagger})$, often referred to as potential, has to be Hermitian and growing sufficiently fast at infinity. To simplify calculations, we split the potential into the Gaussian and the residual part. The Gaussian part can be conveniently parameterized with $\sigma>0$ and $\tau\in [-1,1]$~\cite{FyodorovElliptic}
\begin{equation}
V_G(X,X^{\dagger})=\frac{1}{\sigma^2(1-\tau^2)}\left(XX^{\dagger}-\frac{\tau}{2}\left(X^2+(X^{\dagger})^2\right)\right). \label{eq:GaussianPotential}
\end{equation}
Averages with respect to the Gaussian potential by the virtue of Wick's theorem reduce to products of pairwise expectations, called propagators
\begin{equation}
\<X_{ab}X_{cd}\>_G=\frac{\sigma^2 \tau}{N}\delta_{ad}\delta_{bc},\qquad \<X_{ab}X^{\dagger}_{cd}\>_G=\frac{\sigma^2}{N}\delta_{ad}\delta_{bc}. \label{eq:propagators}
\end{equation}
The exponent of the residual part of the potential is expanded into series, which produces additional terms, called vertices, to be averaged with respect to the Gaussian distribution. To cope with the multitude of terms, we represent them as diagrams (see Table~\ref{Tab:DiagRules} for an overview). This introduces a natural hierarchy of diagrams according to their scaling with the size of the matrix. The dominant contribution, which is of the order of 1, comes from planar diagrams (see Fig.~\ref{Fig:PlanarDiagrams}). The subleading corrections can be classified by the genus of the surface at which they can be drawn without the intersection~\cite{EynardSurfaces}.

\subsection{Moment expansion of the quaternionic resolvent}

\begin{table}\begin{center}
\begin{tabular}{|c|c|c||c|c|c|} \hline
propagator & $\<\cX_{\alpha\beta,ij}\cX_{\mu\nu,jk}\>_G$ & \parbox[c]{2cm}{\includegraphics[width=2cm]{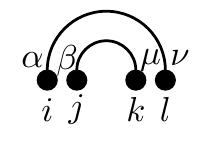}} &
\begin{minipage}{1.5cm}\begin{center}
Green's \\ function \end{center}\end{minipage}  & $G_{\alpha\beta}=\frac{1}{N}\btr \cG$ & \parbox[c]{2cm}{\includegraphics[width=2cm]{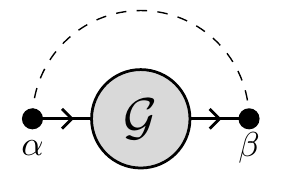}}  \\ \hline
\begin{minipage}{1.5cm} \begin{center} horizontal \\ line \end{center}\end{minipage} & $(Q^{-1})_{\alpha\beta}\delta_{ij}$ & \parbox[c]{1.5cm}{\includegraphics[width=1.5cm]{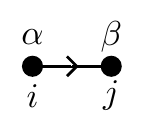}} & 
 vertex & $Ng_3 X_{ij}^{\alpha} X_{jk}^{\beta} X_{ki}^{\gamma}$ & \parbox[c]{1.5cm}{\includegraphics[width=1.5cm]{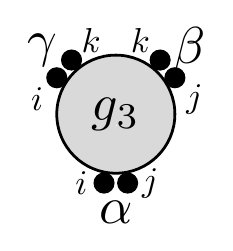}} \\ \hline
 resolvent & $\cG=\<(\cQ-\cX)^{-1}_{\alpha\beta,ij}\>$ & \parbox[c]{2cm}{\includegraphics[width=2cm]{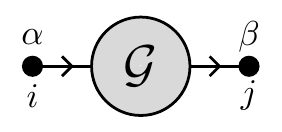}} & 
 cumulant & $\<X^{\alpha}_{ij}X^{\beta}_{jk}X^{\gamma}_{ki}\>_c$ & \parbox[c]{2.5cm}{\includegraphics[width=2.5cm]{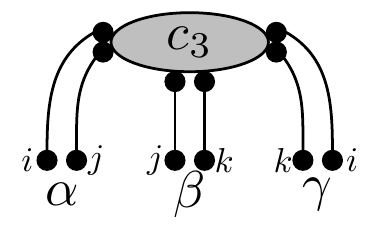}}
 \\ \hline
\end{tabular}
\end{center}
\caption{Diagrammatic representation of the basic expressions in the moment expansion of the resolvent. The propagator represents the averages with respect to the Gaussian potential~\eqref{eq:propagators}. An exemplary vertex is drawn for the theory which contains the cubic interaction $Ng_3\tr X^{\alpha}X^{\beta}X^{\gamma}$ in the potential. A cumulant (dressed vertex) represents a sum over all connected diagrams connected to the baseline. Its structure in matrix  indices (Latin letters) is the same as that of the vertex, because the propagators are the Kronecker deltas in this indices. The dashed line without arrows represent summation over Latin indices only.
\label{Tab:DiagRules}}
\end{table}

To calculate the average of the quaternionic resolvent,  we write it as $\cG=\<\left(\idm -\cQ^{-1}\cX\right)^{-1}\>\cQ^{-1}$ and expand it into the geometric series
\begin{equation}
\cG=\cQ^{-1}+\<\cQ^{-1}\cX\cQ^{-1}\>+\<\cQ^{-1}\cX\cQ^{-1}\cX\cQ^{-1}\>+\ldots, \label{eq:GeometricSeries}
\end{equation}
and perform averages in the large $N$ limit, as described in the previous section. The expansion is valid, provided that $||\cQ^{-1}\cX||<1$, thus for $z$ inside the spectrum of $X$, we need to keep $w$ finite. If the spectrum is bounded, one can always find sufficiently large $w$, so that this series is absolutely convergent.  For the calculations with $z$ outside the spectrum one can safely set $w=0$.

It is convenient to introduce a notation, which incorporates the block structure of the duplicated matrices. We therefore endow each matrix with two sets of indices, writing for example $\cG_{\alpha\beta,ij}$. The first two Greek indices, which we also refer to as quaternionic indices, enumerate blocks and take values $1$ and $\ob$. The Latin ones, running from 1 to $N$ enumerate matrices within each block. The space described by the Latin indices we call simply the matrix space. The block trace operation can be represented as a partial trace over the matrix space $G(Q)_{\alpha\beta}=\frac{1}{N}\sum_{i=1}^{N}\cG_{\alpha\beta,ii}$ (see also  Table~\ref{Tab:DiagRules}). Due to the fact, that the propagators are expressed in terms of Kronecker deltas, all averaged expressions have trivial matrix structure, e.g. $\cG=G\otimes \idm$, but we need this notation for the next section.

\begin{figure}
\includegraphics[width=0.99\textwidth]{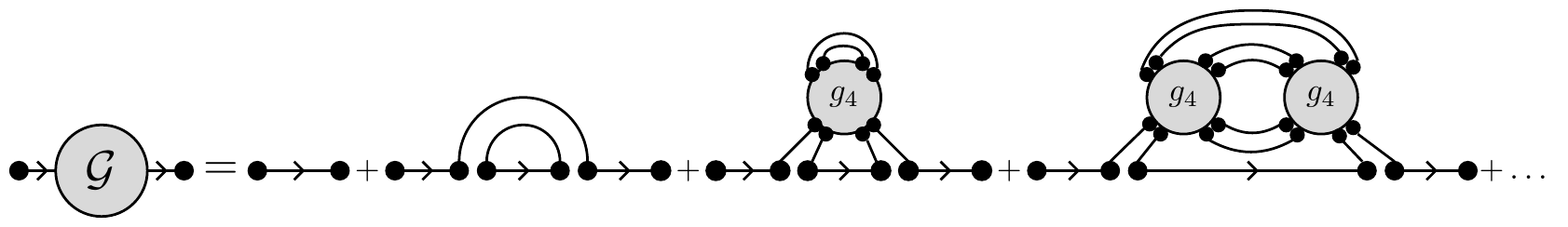}
\caption{ Some exemplary planar diagrams in a model with a quartic term $g_4 X^4$  contributing to the Green's function. All diagrams (except for the first) are 1PI.
\label{Fig:GreenDiagrams}
}
\end{figure}

\begin{figure}
\begin{center}
\includegraphics[width=0.7\textwidth]{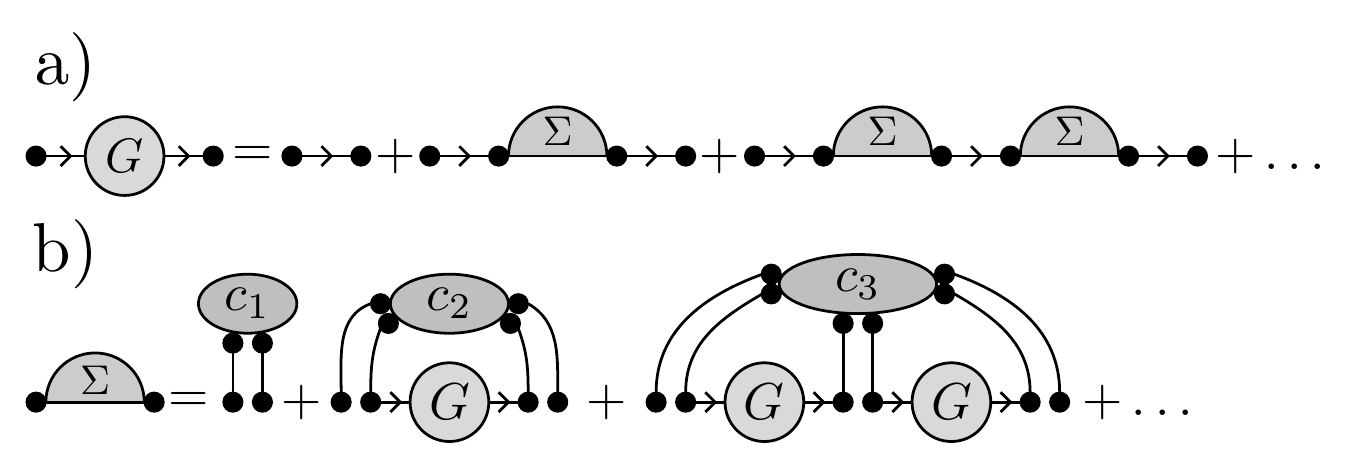}
\end{center}
\caption{ a) First Schwinger-Dyson equation. Diagrams contributing to the Green's function can be divided into one-particle irreducible (1PI) and the ones composed of 1PI connected by a horizontal line (corresponding to $\cQ^{-1}$). 
b) Second Schwinger-Dyson equation. Any 1PI planar diagram can be represented as a certain connected subdiagram attached to the baseline (horizontal line from the graphical representation of the expansion \eqref{eq:GeometricSeries}) via $k$ propagators (this is the $k$-th cumulant). The diagrams between the legs of the cumulant can be of any type, which are in turn encoded in the Green's function. Since all cumulants are encoded in their generating function - the quaternionic $R$-transform \eqref{eq:Rquaternionic}, this relation leads to the equation $\Sigma(Q)=R(G(Q))$.
\label{Fig:SD}
}
\end{figure}

Among all diagrams contributing to $\cG$  (see Fig.~\ref{Fig:GreenDiagrams} for an example) we distinguish a class of one-line irreducible diagrams (1LI), i.e. the ones that cannot be split into two parts, connected only through $\cQ^{-1}$. Let us denote as $\Sigma$ a sum of all 1LI diagrams. This is a building block of the quaternionic resolvent, since any diagram can be decomposed into 1LI subdiagrams connected through $\cQ^{-1}$. Having the absolute convergence of the series, we rearrange terms, obtaining the Schwinger-Dyson equation (here we restrict it only to the quaternionic part)
\begin{equation}
G(Q)=Q^{-1}+Q^{-1}\Sigma(Q)Q^{-1}+Q^{-1}\Sigma(Q)Q^{-1}\Sigma(Q)Q^{-1}+\ldots,
\end{equation}
 presented also diagrammatically in Fig.~\ref{Fig:SD}a). This is a geometric series, which can be summed and written in a closed form
\begin{equation}
G(Q)=\left(Q-\Sigma(Q)\right)^{-1}.\label{eq:SchwDys1}
\end{equation}

\subsection{Quaternionic R-transform}

To find $G$, one needs to relate $\Sigma$ to $G$. To this end, let us consider diagrams contributing to averages of traced strings of $X$'s and $X^{\dagger}$'s such that all $X$'s and $X^{\dagger}$'s are connected with each other. Their sum we call a cumulant (in field theory language it is known as a dressed vertex) and endow the respective average with a subscript $c$. We adopt a convenient notation for cumulants in which $\dagger$ is associated with the $\ob$ index and, trivially, lack of conjugation with $1$. We also encode the length of the string. An example reads
\begin{equation}
c^{(k)}_{\alpha_1\alpha_2\ldots \alpha_k}=\<\frac{1}{N}\tr X^{\alpha_1}X^{\alpha_2}\ldots X^{\alpha_k}\>_c.
\end{equation}
We also introduce the $R$-transform, which is a $2\times 2$ matrix, defined through its expansion for small arguments
\begin{equation}
R_{\alpha\beta}=c^{(1)}_{\alpha}\delta_{\alpha\beta}+c^{(2)}_{\alpha\beta}Q_{\alpha\beta}+\sum_{\mu\in\{1,\ob\} }c^{(3)}_{\alpha\mu\beta}Q_{\alpha\mu}Q_{\mu\beta}+\sum_{\mu,\nu\in\{1,\ob\} }c^{(4)}_{\alpha\mu\nu\beta}Q_{\alpha\mu}Q_{\mu\nu}Q_{\nu\beta}+\ldots \label{eq:Rquaternionic}
\end{equation}
This definition written in terms of indices may not seem to be intuitive, but in the matrix notation takes a clearer form
\begin{equation}
R(Q)\otimes \idm= \<\cX\>_c+\<\cX\cQ\cX\>_c+\<\cX\cQ\cX\cQ\cX\>_c+\ldots,
\end{equation}
which is the counterpart of \eqref{eq:GeometricSeries}. The matrix $R$ is also a quaternion, so it is dubbed the quaternionic $R$-transform. $Q$ is always associated with two consecutive indices in the cumulant and can be thought of as a transfer matrix. It is crucial for encoding all cumulants in the $R$-transform that matrices $\cX$ and $\cQ$ do not commute.

Consider now any 1LI diagram. Due to its irreducibility it can be depicted as a certain cumulant connecting the first and last $\cX$ and possibly some others in between. The subdiagrams disconnected from the cumulant can be in any form, which is already encoded in the quaternionic Green's function. This allows us to write the second Schwinger-Dyson equation relating $\Sigma$ and $G$ via the quaternionic $R$-transform  (see also Fig.~\ref{Fig:SD}b)) 
\begin{equation}
\Sigma(Q)=R(G(Q)). \label{eq:SchwDys2}
\end{equation}
The knowledge of all cumulants allows us to solve the matrix model, since equations \eqref{eq:SchwDys1} and \eqref{eq:SchwDys2} can be brought to a single $2\times 2$ matrix equation
\begin{equation}
R(G(Q))+G(Q)^{-1}=Q. \label{eq:ForGreensFunction}
\end{equation}
 Once the averaging with respect to the ensemble was taken at the level of diagrams, we can safely remove the regularization and solve the above algebraic equation, setting first $w=0$. We refer to~\cite{HLdiag,nowak2017lagged} for more detailed calculations in the diagrammatic formalism.
 
 The construction presented in this section has been recently rigorously formalized in the framework of free probability~\cite{BSS}.

\section{2-point eigenvector correlation function \label{Sec:Main}}

\subsection{Preliminaries}

In order to investigate the 2-point eigenvector correlation function associated with the off-diagonal overlap, we follow the paradigm outlined in the previous section for calculations of Green's function. A naive approach, i.e. calculation of $\mathfrak{h}(z_1,\zb_2)=\<\frac{1}{N}\tr (z_1\idm-X)^{-1}(\zb_2\idm-X^{\dagger})^{-1}\>$, yields the result which is correct only outside of the spectrum of $X$, which we refer to as the holomorphic solution. Inside the spectrum, we regularize each resolvent, using the rule
\begin{equation}
(z\idm-X)^{-1}\rightarrow (\zb\idm-X^{\dagger})M(z,w)^{-1},
\end{equation}
where $M(z,w)=(z\idm-X)(\zb\idm-X^{\dagger})+|w|^2\idm$. We shall therefore study
\begin{equation}
h(z_1,w_1,z_2,w_2)=\<\frac{1}{N}\tr (\zb_1\idm-X^{\dagger})M(z_1,w_1)^{-1}M(z_2,w_2)^{-1}(z_2\idm-X)\>. \label{eq:RegularizedH}
\end{equation}

The weighted density is therefore given by
\begin{equation}
D(z_1,z_2)=\lim_{|w_1|,|w_2|\to 0} \frac{1}{\pi^2} \partial_{\zb_1}\partial_{z_2} h(z_1,w_1,z_2,w_2).
\end{equation}

In this paper we will calculate $h$ by diagrammatic $1/N$ expansion in the planar limit. The singular part of $D(z_1,z_2)$ containing the Dirac delta is not accessible in perturbative calculations, so we get
\begin{equation}
O_2(z_1,z_2)=\lim_{|w_1|,|w_2|\to 0} \frac{1}{\pi^2} \partial_{\zb_1}\partial_{z_2} h(z_1,w_1,z_2,w_2). \label{eq:OFormula}
\end{equation}

There exists a class of matrices which despite not being Hermitian have a real spectrum. A simple example is the product of two Hermitian matrices $A,B$, one of which (let us say $A$) is positive definite. The resulting matrix is not Hermitian, but isospectral to $A^{1/2}BA^{1/2}$, which must have real eigenvalues. The eigenvectors of $AB$ are not orthogonal, which makes $O_2$ non-trivial. The realness of the spectrum facilitates calculations, as the knowledge of the traced resolvent is sufficient. By the virtue of the Sochocki-Plemelj formula valid for real $x$ we can write
\begin{equation}
2\pi i\delta(x)=\lim_{\epsilon\to 0}\left(\frac{1}{x-i\epsilon}-\frac{1}{x+i\epsilon}\right),
\end{equation}
and the two-point function can be calculated from the holomorphic function via
\begin{equation}
O_2(x,y)=\frac{-1}{4\pi^2}\left(\mathfrak{h}(+,+)-\mathfrak{h}(+,-)-\mathfrak{h}(-,+)+\mathfrak{h}(-,-)\right), \label{eq:CorrelationPTmodel}
\end{equation}
where 
\begin{equation}
\mathfrak{h}(\pm,\pm)=\lim_{\epsilon_1,\epsilon_2\to 0}\mathfrak{h}(x\pm i\epsilon_1,y\pm i\epsilon_2) \label{eq:PTcorrelatorRule}
\end{equation}
and signs are uncorrelated.

\subsection{Linearization}
The expression for the regularized product of resolvents~\eqref{eq:RegularizedH} contains two quadratic nonlinearities. We overcome this difficulty, by using $2N\times 2N$ matrices $\cQ=Q\otimes \idm$, $\cP=P\otimes \idm$ and $\cX$, where
\begin{equation}
Q=\left(\begin{array}{cc}
z_1  & i\wb_1  \\
i w_1  & \zb_1 
\end{array}\right),\quad
P=\left(\begin{array}{cc}
z_2  & i\wb_2  \\
i w_2  & \zb_2 
\end{array}\right), \quad 
\cX=\left(\begin{array}{cc}
X & 0 \\
0 & X^{\dagger}
\end{array}\right).
\end{equation}
 As a natural generalization of the quaternionic resolvent to two-point functions, we define the average of the Kronecker product of two quaternionic resolvents
 \begin{equation}
 \cH=\<(\cQ-\cX)^{-1}\otimes (\cP^T-\cX^T)^{-1}\>.
 \end{equation}
Such an object is quite unusual in Random Matrix Theory. A similar construction was used by Brezin and Zee for the calculation of the connected 2-point density in Hermitian models~\cite{brezinzee1995TwoPoint}, but there one deals only with matrix indices.  To the best of our knowledge the quaternionic approach to two-point functions for non-Hermitian matrices is considered for the first time, thus we will discuss it in more detail.

 $\cH$ is a $4N^2\times 4N^2$ matrix with a very specific block structure. To keep track of it, we endow $\cH$ with 8 indices. The upper ones refer to the first matrix in the Kronecker product, while the lower ones to the second. As in the case of the quaternionic Green's function, Greek indices, taking values in $\{1,\ob\}$, enumerate blocks, while Latin indices ranging in $\{1,\ldots,N\}$ denote elements within each block. In the index notation, its components read (note the transpose of the second matrix)
 \begin{equation}
 \cH^{\alpha\beta,ij}_{\mu\nu,kl}=\<\left(\cQ-\cX\right)^{-1}_{\alpha\beta,ij} \left(\cP-\cX\right)^{-1}_{\nu\mu,lk}   \>.
 \end{equation}
 
With the same assumptions as for one-point functions, the resolvents are then expanded into the power series
\begin{equation}
\cH=\<\left(\cQ^{-1}+\cQ^{-1}\cX\cQ^{-1}+\ldots\right)\otimes \left( \cP^{-1}+\cP^{-1}\cX\cP^{-1}+\ldots\right)^T\>,
\end{equation}
and taking the expectation produces diagrams. The flow of Latin (matrix) indices in the diagrams follows the lines in the double line notation. The propagators are symmetric, thus the direction does not matter. The flow of quaternionic  (Greek) indices is governed by their order in the expansion of the resolvent. Since the quaternion matrices $Q$ and $P$ are not symmetric, the direction  of the line representing $\cQ^{-1}$ matters and is depicted by an arrow. We draw diagrams in such a way that the terms in the expansion of the resolvents are in two rows, hereafter called baselines, with the first resolvent above. The quaternionic indices flow from left to right in the upper baseline and in the opposite direction below.

\begin{figure}
\begin{center}
\includegraphics[height=4cm]{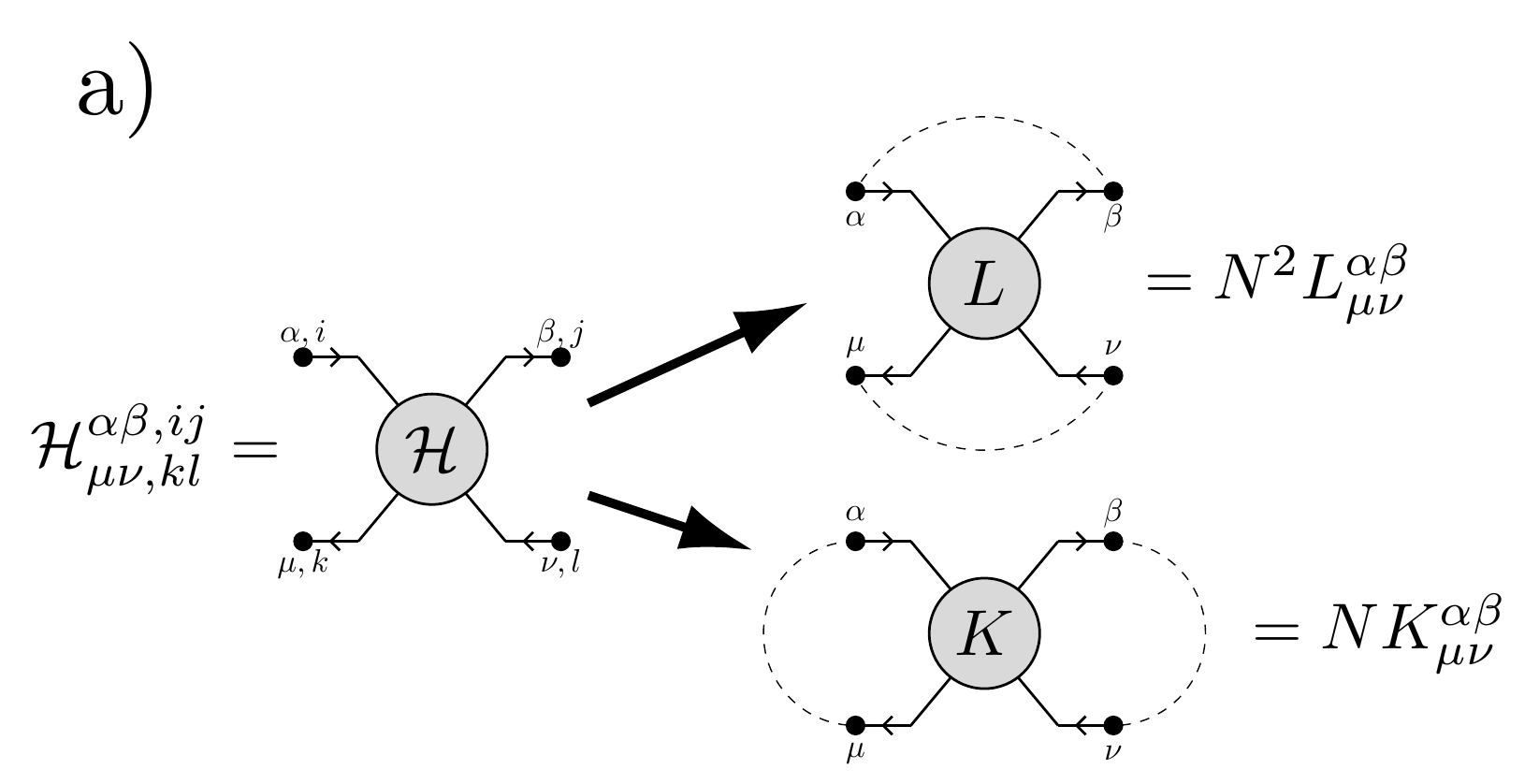}  \includegraphics[height=4cm]{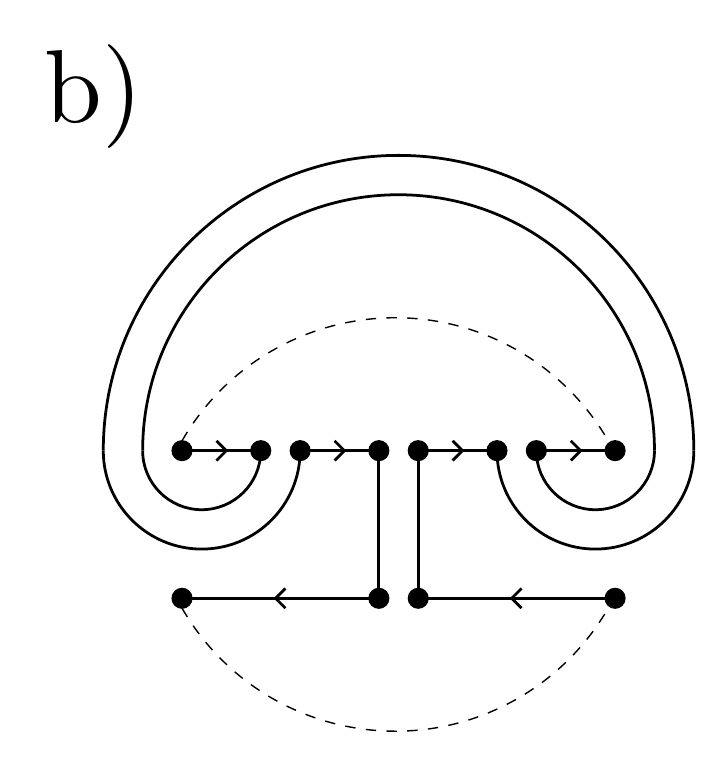} \includegraphics[height=4cm]{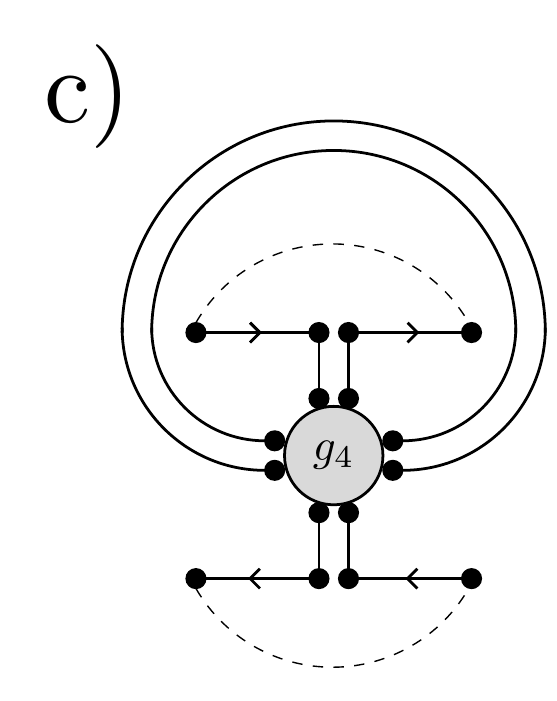}
\end{center}
\caption{ a) Possible contractions of matrix indices (dashed lines) of the Kronecker product of two quaternionic resolvents. The way  one contracts  indices determines the class dominant planar diagrams, which are drawn between dashed lines and the horizontal baselines. The upper choice corresponds to a class of double-trace two-point functions, see \eqref{eq:logdet2}, while the lower possibility leads a single-trace two point function encoding correlations of eigenvectors. Diagrams contributing to $L$ are of wheel type~\cite{brezinzee1995TwoPoint,lukaszewski2008diagrammatic} and $K$ is given as a sum of ladder diagrams. b) An example of a diagram which contributes to $L$ but is subleading in the calculation of $K$. c) An example of a diagram appearing during the calculation of $L$, which despite its planarity is subleading.
\label{fig:Contraction}
}
\end{figure}

\begin{figure}
 \includegraphics[width=\textwidth]{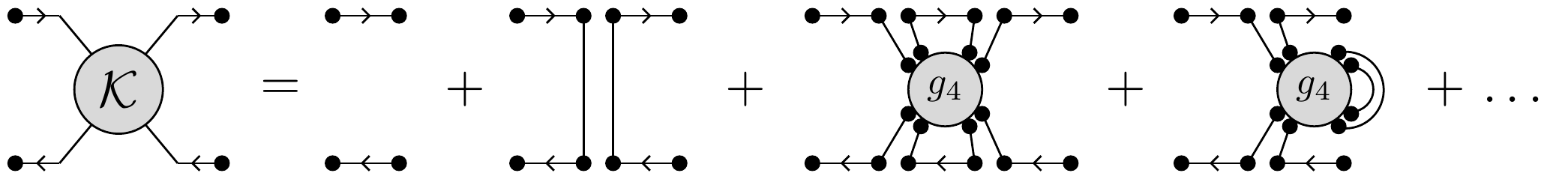}
 \caption{Some exemplary diagrams in a theory with quartic potential contributing to $\cK$.
 \label{Fig:Hexp}
 }
 \end{figure}

There are two ways of contracting matrix indices\footnote{In fact, there are $\frac{4!}{2^2\cdot 2!} =3$ ways, but $\sum_{ij}\cH^{\alpha\beta,ij}_{\mu\nu,ji}$ leads to the same diagrams as $K^{\beta\alpha}_{\mu\nu}$.}, thus we define two functions
\begin{equation}
K^{\alpha\beta}_{\mu\nu}=\frac{1}{N}\sum_{i,j=1}^{N}\cH^{\alpha\beta,ij}_{\mu\nu,ij},\qquad L ^{\alpha\beta}_{\mu\nu}=\frac{1}{N^2}\sum_{i,j=1}^{N} \cH^{\alpha\beta,ii}_{\mu\nu,jj},
\end{equation} 
which correspond to contractions presented in Fig.~\ref{fig:Contraction}a). It will become clear later that $K$ encodes correlations of eigenvectors and $L$ of eigenvalues. These two possible contractions define two different classes of planar diagrams. The admissible diagrams have to be drawn in the region of the plane bounded by baselines and dashed lines depicting contractions. The diagrams contributing to $K$ are of the ladder type (see Fig.~\ref{Fig:Hexp}), while the class of planar diagrams contributing to $L$, termed wheel diagrams, is broader, as it admits for circumventing one of the baselines if the points on the baseline are connected through propagators and vertices, see Fig.~\ref{fig:Contraction}b). Not all planar diagrams contribute equally to $L$. Diagrams in which a propagator connects two sides of a vertex and encircles a baseline is subleading, see Fig.~\ref{fig:Contraction}c). In this section we concentrate on the ladder diagrams.

\subsection{Ladder diagrams } 
\label{sec:Diagrams}

In this section we are interested in the calculation of $K$. The contraction of matrix indices in $\cH$, which leads to $K$, is in fact a summation of all $N^4$ elements within each $4\times 4$ block. To make the notation of Greek indices even more explicit, we write the entries of $K$
\begin{equation}
K=\left(\begin{array}{cccc}
\displaystyle K^{11}_{11} & K^{11}_{1\ob} & K^{1\ob}_{11} & K^{1\ob}_{1\ob} \\
\displaystyle K^{11}_{\ob 1} & K^{11}_{\ob\ob} &  K^{1\ob}_{\ob 1} & K^{1\ob}_{\ob\ob} \\
\displaystyle K^{\ob 1}_{11}  &  K^{\ob 1}_{1\ob} & K^{\ob\ob}_{11} & K^{\ob\ob}_{1\ob} \\
\displaystyle K^{\ob 1}_{\ob 1}  &  K^{\ob 1}_{\ob\ob} & K^{\ob\ob}_{\ob 1} & K^{\ob\ob}_{\ob\ob}
\end{array}\right). \label{eq:ComponentsEnumeration}
\end{equation}
 An important consequence of this construction is that the $K^{11}_{\ob\ob}$ element is exactly the desired function $h$ \eqref{eq:RegularizedH} for the calculation of the eigenvector correlation function.

\begin{figure}
\begin{center}
\includegraphics[width=\textwidth]{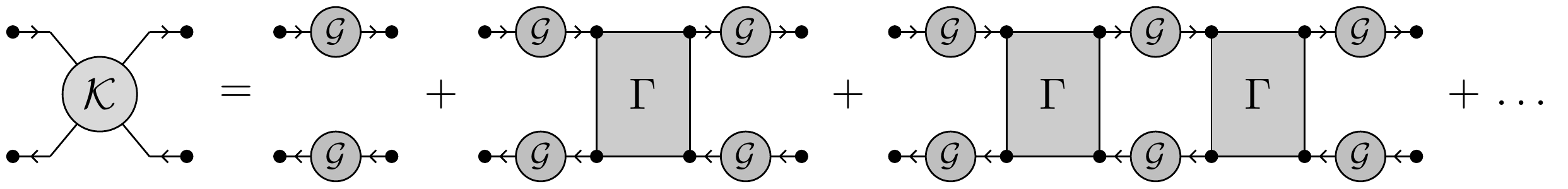}
\end{center}
\caption{ The general structure of planar ladder diagrams contributing to $\cK$.
\label{fig:LadderDiagrams}
}
\end{figure}

\begin{figure}
\begin{center}
\includegraphics[height=2cm]{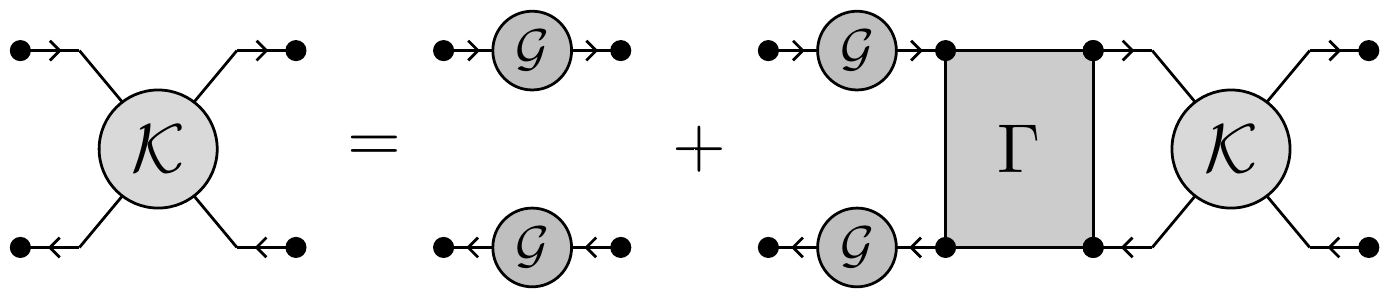}
\end{center}
\caption{ Matrix Bethe-Salpeter equation~\eqref{eq:BetheSalpeter}. 
\label{fig:BetheSalpeter}
}
\end{figure}

Let us define $\cK^{\alpha\beta,ij}_{\mu\nu,kl}$ the sum of all ladder diagrams contributing to $K$ (before we contract indices).  A vertex can connect two points on a baseline (a side rail of the ladder), dressing the part of the rail. There are also vertices connecting two baselines, which give rise to the rungs of the ladder. If we denote $\Gamma^{\alpha\beta,ij}_{\mu\nu,kl}$ a sum of  all connected subdiagrams which connect two rails, one can express $\cK$ in terms of $\Gamma$ as a geometric series, presented in Fig.~\ref{fig:LadderDiagrams}, which can be written in a closed form (a sum over repeating indices is implicit)
\begin{equation}
\cK^{\alpha\beta,ab}_{\mu\nu,cd}=G^{\alpha\beta}G_{\mu\nu}\delta^{ab}\delta_{cd}+G^{\alpha\gamma}G_{\mu\rho}\delta^{ai}\delta_{cj}\Gamma^{\gamma\epsilon,ik}_{\rho\sigma,jl}\cK^{\epsilon\beta,kb}_{\sigma\nu,ld}. \label{eq:BetheSalpeter}
\end{equation}
This relation, shown diagramatically in Fig.~\ref{fig:BetheSalpeter} and known as the matrix Bethe-Salpeter equation, is the counterpart of the Schwinger-Dyson equation for the two-point function, with $\Gamma$ the counterpart of the self-energy.

\begin{figure}
\begin{center} \includegraphics[width=0.3\textwidth]{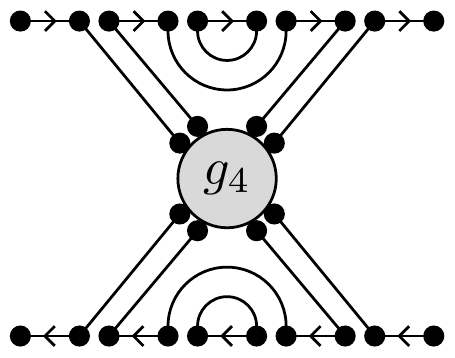} \end{center}
 \caption{
An example of a diagram contributing to $\Gamma$. It contributes to the second-to-last diagram in Fig.~\ref{fig:GammaCumulants}. Since the matrix indices follow the solid lines and propagators are given by Kronecker deltas, $\Gamma^{\alpha\beta,ij}_{\mu\nu,kl}=\frac{1}{N} B^{\alpha\beta}_{\mu\nu} \delta^{i}_k \delta^j_l$, allowing for the calculation of $K$. 
\label{Fig:GammaExample}
 }
 \end{figure}

A direct analysis of planar diagrams yields $\Gamma^{\alpha\beta,ij}_{\mu\nu,kl}=\frac{1}{N} B^{\alpha\beta}_{\mu\nu} \delta^{i}_k \delta^j_l$, where $B$ is of order 1, see Fig.~\ref{Fig:GammaExample}.
Using the matrix structure of $\Gamma$, we trace out the matrix indices and find the equation for $K$, which in the matrix notation reads 
\begin{equation}
K(Q,P)=G(Q)\otimes G^{T}(P)+\left[G(Q)\otimes G^{T} (P) \right]  B(Q,P) K(Q,P). \label{eq:Master1}
\end{equation}

\begin{figure}
\begin{center}
\includegraphics[width=\textwidth]{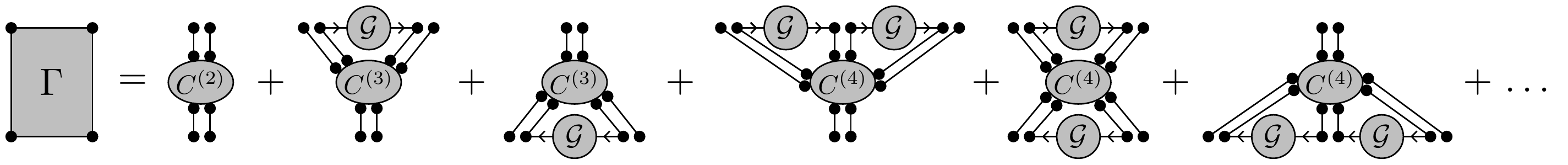}
\end{center}
\caption{ $\Gamma$ given by the planar cumulants.
\label{fig:GammaCumulants} }
\end{figure}

We now turn our attention to the rungs. Any diagram contributing to $\Gamma$ can be decomposed as a certain cumulant of length $n\geq 2$, the first $k$ legs of which are attached to the upper rail, while the last legs are connected to the lower rail. The part of the rail between the legs of the cumulant gets dressed to the quaternionic Green's function $G(Q)$ above and $G(P)$ below. The space between $k$-th and $(k+1)$-th legs is left unfilled, because the quaternonic indices at the end of rails are not contracted. This decomposition of $\Gamma$ is depicted in Fig.~\ref{fig:GammaCumulants}.
As $\Gamma$ is completely determined by the planar cumulants, $B^{\alpha\beta}_{\mu\nu}$ can be calculated from the quaternionic $R$ transform \eqref{eq:Rquaternionic}. The rule is simple and goes as follows. 

Consider the expansion of $R_{\alpha\mu}$ in $Q$ \eqref{eq:Rquaternionic} and differentiate it with respect to $Q_{\beta\nu}$. As a result for some $0<k<n-1$ the $k$-th quaternion $Q_{\mu_k \mu_{k+1}}$ will be replaced by $\delta_{\mu_k \beta}\delta_{\nu\mu_{k+1}}$. Then all $Q_{\mu_l\mu_{l+1}}$'s from the lhs of the removed $Q$ (i.e. for $l<k$) are replaced by $G^{\mu_l\mu_{l+1}}(Q)$ and all $Q_{\mu_l\mu_{l+1}}$ on the right ($l>k$) by $G^{T}_{\mu_l\mu_{l+1}}(P)$. Then the sum over all possible positions (i.e. $k$'s), where $\cQ$ has been removed, is performed.

$B$ can be therefore expressed in terms of cumulants as a power series 
\begin{equation}
B^{\alpha\beta}_{\mu\nu}(Q,P)=\sum_{k,l=1}^{\infty} \sum_{\substack{\sigma_1,\ldots,\sigma_{k} \\ \rho_1,\ldots,\rho_{l}} } \delta^{\alpha\sigma_1}\delta^{\beta\sigma_k}\delta_{\mu\rho_l}\delta_{\nu\rho_1} c^{(k+l)}_{\sigma_1\ldots \sigma_{k}\rho_1\ldots\rho_{l}}G^{\sigma_1\sigma_2}(Q)\ldots G^{\sigma_{k-1}\sigma_k}(Q)G_{\rho_1\rho_2}(P)\ldots G_{\rho_{l-1}\rho_l}(P), \label{eq:Apowerseries}
\end{equation}
where all $\sigma_i$ and $\rho_j$ take values in $\{1,\ob\}$ and for $k=1$ or $l=1$ $G_{\sigma_k\sigma_{k+1}}$ reduces to Kronecker delta. An application of this procedure to the quantum scattering ensemble is presented in Appendix \ref{sec:QuantumScattering}.

 We remark that the additivity of the quaternionic $R$-transform under the addition of unitarily invariant non-Hermitian matrices implies additivity of $B$.

\subsection{Traced product of resolvents}

In the holomorphic domain outside the spectrum the situation simplifies considerably, because one can set $|w|\to 0$ at the very beginning of calculations. Green's function is then diagonal, $G(z,\zb)=\diag(\mathfrak{g}(z),\bar{\mathfrak{g}}(\zb))$, where  $\mathfrak{g}(z)=\<\frac{1}{N}\tr (z\idm-X)^{-1}\>$ and $\bar{\mathfrak{g}}(\zb)=\<\frac{1}{N}\tr (\zb\idm-X^{\dagger})^{-1}\>$. Due to such a structure, $B$ is also diagonal with components 
\begin{equation}
B^{\alpha\beta}_{\mu\nu}=\delta^{\alpha\beta}\delta_{\mu\nu}\sum_{k,l=1}^{\infty}c^{(k+l)}_{\underbrace{\alpha\ldots\alpha}_{k} \underbrace{\mu \ldots \mu}_{l}} \left(\mathfrak{g}_{\alpha}(z_1)\right)^{k-1}\left(\mathfrak{g}_{\mu}(z_2)\right)^{l-1}, \label{eq:Aholom}
\end{equation}
where we assume the standard convention $\mathfrak{g}_{1}(z)=\mathfrak{g}(z)$ and $\mathfrak{g}_{\ob}(z)=\bar{\mathfrak{g}}(\zb)$. A matrix equation~\eqref{eq:Master1} splits into decoupled scalar equations with the explicit solution for the component of our interest
\begin{equation}
K_{\ob\ob}^{11}=\frac{\mathfrak{g}(z_1)\bar{\mathfrak{g}}(\zb_2)}{1-\mathfrak{g}(z_1)\bar{\mathfrak{g}}(\zb_2) B^{11}_{\ob\ob}}. \label{eq:twopoint}
\end{equation}
The desired component of $B$ obtained from~\eqref{eq:Aholom} reads
\begin{equation}
B^{11}_{\ob\ob}=\sum_{k,l=1}^{\infty}c^{(k+l)}_{\underbrace{1\ldots 1}_{k} \underbrace{\ob \ldots \ob}_{l}} \left(\mathfrak{g}(z_1)\right)^{k-1}\left(\bar{\mathfrak{g}}(\zb_2)\right)^{l-1}, \label{eq:A11}
\end{equation}

 Despite the fact that the mapping between cumulants and the $R$-transform is not one to one~\cite{HLdiag}, some cumulants can be uniquely determined from the knowledge of $R(Q)$.  The cumulants $c_{\underbrace{1\ldots 1}_{k}\underbrace{\ob \ldots \ob}_{n-k}}^{(n)}$ are the coefficients at $Q_{11}^{k-1}Q_{1\ob}Q_{\ob\ob}^{n-k-1}$ in the expansion of $R_{1\ob}(Q)$. One can easily see that there are no other cumulants contributing to this term. 
 
 All cumulants contributing to $R_{1\ob}$ have at least one $X^{\dagger}$ following $X$ in the string, therefore $R_{1\ob}$ is divisible by $Q_{1\ob}$. Let us define $\tilde{R}_{1\ob}=R_{1\ob}/Q_{1\ob}$, which is regular at 0. The considered cumulants are the only ones in which $X$ is followed by $X^{\dagger}$ exactly once. To exclude all other possibilities in the expansion of $\tilde{R}_{1\ob}$, we set $Q_{1\ob}=0=Q_{\ob 1}$ in $\tilde{R}_{1\ob}(Q)$. To reproduce \eqref{eq:A11} from $\tilde{R}_{1\ob}$ one also needs to replace $Q_{11}$ by $\mathfrak{g}(z_1)$ and $Q_{\ob\ob}$ by $\bar{\mathfrak{g}}(\zb_2)$. Finally,
\begin{equation}
B^{11}_{\ob\ob}=\tilde{R}_{1\ob}\left(\diag(\mathfrak{g}(z_1),\bar{\mathfrak{g}}(\zb_2))\right).
\end{equation}

\subsection{Biunitarily invariant ensembles}
\label{sec:Rdiag}
In this subsection we consider a class of ensembles, the pdf of which is invariant under  multiplication by two unitary matrices, i.e. $P(X)=P(UXV^{\dagger})$. In the large $N$ limit the spectral density, which is rotationally invariant, is supported on either a disc or an annulus, a phenomenon termed `the single ring theorem'~\cite{FeinbergZeeSingleRing,GuionnetKrishnapurZeitouni}. The enhanced symmetry allows one to relate the distribution of eigenvalues and singular values both in the $N\to\infty$ limit~\cite{HaagerupLarsen} and for finite $N$~\cite{KieburgKosters}.  More precisely, the radial cumulative distribution function $F(r)=2\pi \int_0^r\rho(s)sds$ is given by the solution of the simple functional equation $S_{XX^{\dagger}}(F(r)-1)=\frac{1}{r^2}$, where $S_{XX^{\dagger}}$ is the Voiculescu $S$-transform of the density of squared singular values~\cite{HaagerupLarsen}. Recently, this result was extended to the one-point eigenvector correlation function, which is determined solely by $F$~\cite{CondNum,HLdiag}
\begin{equation}
O_1(r)=\frac{F(r)(1-F(r))}{\pi r^2}.
\end{equation}

Such simple results in the large $N$ limit are possible because of the exceptionally simple structure of cumulants. The only non-zero planar cumulants are the alternating ones~\cite{HLdiag}, $\alpha_n=c^{(2n)}_{1\ob\ldots 1\ob}=c^{(2n)}_{\ob 1\ldots \ob 1}$. They can be encoded in a function of a single scalar variable $A(x)=\sum_{k=1}^{\infty} \alpha_n x^{n-1}$, called the determining sequence~\cite{DetSeq}. Due to this, only four components  of $B$ (out of 16) do not vanish. These are $B^{11}_{\ob\ob}=B^{\ob\ob}_{11}$, $B^{1\ob}_{\ob 1}$, $B^{\ob 1}_{1\ob}$.

A direct application of formula \eqref{eq:Apowerseries} leads to
\begin{align}
B^{11}_{\ob\ob} &  =\sum_{k,l=1}^{\infty} \alpha_{k+l-1} \left[G_{1\ob}(Q)G_{\ob 1}(Q)\right]^{k-1} \left[ G_{1\ob}(P) G_{\ob 1}(P)\right]^{l-1}\\
B^{1\ob}_{\ob 1}&  =G_{1\ob}(Q)G_{1\ob}(P)\sum_{k,l=1}^{\infty} \alpha_{k+l}\left[G_{1\ob} (Q) G_{\ob 1}(Q)\right]^{k-1}\left[G_{1\ob} (P) G_{\ob 1}(P)\right]^{l-1} \\
B^{\ob 1}_{1\ob}&  =G_{\ob 1}(Q)G_{\ob 1}(P)\sum_{k,l=1}^{\infty}\alpha_{k+l} \left[G_{1\ob}(Q)G_{\ob 1}(Q)\right]^{k-1} \left[G_{1\ob}(P)G_{\ob 1}(P) \right]^{l-1}
\end{align}

The components of $B$ can be expressed through auxiliary  functions
\begin{align}
B^{11}_{\ob\ob}=B^{\ob\ob}_{11}=S\left(G_{1\ob}(Q)G_{\ob 1}(Q),G_{1\ob}(P)G_{\ob 1}(P) \right), \\
B^{1\ob}_{\ob 1}=G_{1\ob} (Q) G_{1\ob}(P)T\left(G_{1\ob}(Q)G_{\ob 1}(Q),G_{1\ob}(P)G_{\ob 1}(P) \right), \\
B^{\ob 1}_{1\ob}=G_{\ob 1}(Q) G_{\ob 1}(P)   T\left(G_{1\ob}(Q)G_{\ob 1}(Q),G_{1\ob}(P)G_{\ob 1}(P) \right),
\end{align}
where 
\begin{eqnarray}
S(x,y)=\sum_{k,l=1}^{\infty}\alpha_{k+l-1} x^{k-1} y^{l-1}=\frac{xA(x)-yA(y)}{x-y}, \\
T(x,y)=\sum_{k,l=1}^{\infty}\alpha_{k+l}x^{k-1}y^{k-1}=\frac{A(x)-A(y)}{x-y},
\end{eqnarray}
with $A$ being the determining sequence.

We remark that the average over the ensemble has been already taken at the level of Feynman diagrams and at this moment, we can safely remove the regularization. There are further simplifications for the biunitarily invariant matrices~\cite{HLdiag}
\begin{equation}
G_{1\ob}G_{\ob 1}A(G_{1\ob}G_{\ob 1})=F(r)-1, \quad G_{1\ob}G_{\ob 1}=-\pi O_1(r). 
\end{equation}

Having calculated $B$ and knowing Green's function, we determine $K^{11}_{\ob\ob}$ from~\eqref{eq:Master1} and, after algebraic manipulations, we get a compact formula for the 2-point eigenvector correlation function from~\eqref{eq:OFormula}

\begin{equation}
O_2(z_1,z_2)=\frac{1}{\pi} \partial_{\zb_1}\partial_{z_2}\frac{\zb_1 (z_1-z_2)O_1(r_1)+z_2(\zb_1-\zb_2) O_1(r_2)}{|z_1-z_2|^2 \left[F(r_1)-F(r_2)\right]}. \label{eq:ORdiagonal}
\end{equation}

 The quaternionic R-transform of biunitarily invariant ensembles takes a remarkably simple form~\cite{HLdiag}, in particular $R_{1\ob}(Q)=Q_{1\ob}A(Q_{1\ob}Q_{\ob 1})$. Moreover, due to the rotational symmetry of the spectrum, $\mathfrak{g}(z)=1/z$. According to \eqref{eq:twopoint}, the traced product of resolvents is given by
\begin{equation}
\mathfrak{h}(z_1,\zb_2)=\frac{1}{z_1\zb_2-A(0)}.
\end{equation}
Interestingly, $A(0)=r_{out}^2$, where $r_{out}$ is the external radius of the spectrum. This result shows a high level of universality, since for any two functions $f,g$ analytic in the spectrum the expectation in the $N \to \infty$ limit
\begin{equation}
\<\frac{1}{N}\tr f(X)g(X^{\dagger})\>=\frac{1}{(2\pi i)^2}\int_{\gamma}dz_1\int_{\bar{\gamma}}d\zb_2 \frac{f(z_1) g(\zb_2)}{z_1\zb_2-r_{out}^2}
\end{equation}
is given by the same formula, irrespectively of the specific biunitarily invariant ensemble. The only parameter -- spectral radius $r_{out}$ -- can be set to 1 by rescaling the matrix. This result, appearing naturally in the language of cumulants, from the point of the spectral decomposition, $X=\sum_k \ket{R_k}\lambda_k\bra{L_k}$, is far from being obvious and may explain the simplicity of formula \eqref{eq:ORdiagonal}.

\section{Examples }
\label{sec:Examples}

\subsection{Elliptic ensemble}
As the first example of application of this formalism, we take the elliptic ensemble. Due to the fact that only the second cumulants do not vanish, the sum in \eqref{eq:Apowerseries} reduces to a single term and $B$ is diagonal, $B_{ell}=\diag(\sigma^2 \tau,\sigma^2,\sigma^2,\sigma^2 \tau)$. However, the equations \eqref{eq:Master1} do not decouple, because  Green's functions are not diagonal in the non-holomorphic regime. Denoting for $j=1,2$
\begin{equation}
G_{j}=\left(\begin{array}{cc}
\frac{\zb_j-z_j\tau}{\sigma^2(1-\tau^2)} & \frac{i}{\sigma^2}\sqrt{1-\frac{|z_j-\zb_j\tau|^2}{\sigma^2(1-\tau^2)}} \\
\frac{i}{\sigma^2}\sqrt{1-\frac{|z_j-\zb_j\tau|^2}{\sigma^2(1-\tau^2)}} & \frac{z_j-\zb_j\tau}{\sigma^2(1-\tau^2)}
\end{array}\right)
\end{equation}
 Green's function of the elliptic ensemble in the non-holomorphic regime (see Appendix~\ref{sec:Elliptic}), we find $K$, solving \eqref{eq:Master1}
\begin{equation}
K=\left(\idm- (G_1\otimes G_2^T)B_{Ell}\right)^{-1}\left(G_1\otimes G_2^T\right).
\end{equation}
Then we focus on the component $K^{11}_{\ob\ob}$ and differentiate it twice, according to \eqref{eq:OFormula}, obtaining 
\begin{equation}
O_2(z_1,z_2)=\frac{1}{\pi^2}\partial_{\zb_1}\partial_{z_2} K^{11}_{\ob\ob}=-\frac{\sigma^2(1-\tau^2)^2-(z_1-\zb_2\tau)(\zb_2-z_1\tau)}{\pi^2\sigma^2(1-\tau^2)|z_1-z_2|^4}. \label{eq:Elliptic2point}
\end{equation}
This result was derived for the first time by Chalker and Mehlig~\cite{mehlig2000correlator}\footnote{\cite[Eq.(94)]{mehlig2000correlator} contains a small misprint in the constant factor, which does not affect validity of any other results therein.}. For the Ginibre Ensemble ($\sigma=1$, $\tau=0$) it reduces to
\begin{equation}
O_2(z_1,z_2)=\frac{-1}{\pi^2} \frac{1-z_1\zb_2}{|z_1-z_2|^4}. \label{eq:Ginibre}
\end{equation}

For completeness, we remark that the holomorphic part of the two point function, calculated from \eqref{eq:twopoint}, reads
\begin{equation}
\mathfrak{h}(z_1,\zb_2)=\frac{4}{-4+\left(z_1+\sqrt{z_1^2-4\sigma^2\tau}\right)\left(\zb_2+\sqrt{\zb_2^2-4\sigma^2\tau}\right)}.
\end{equation}

\subsection{Biunitarily invariant ensembles}
\label{sec:ExamplesBiunitary}
We consider some examples where the two-point function can be easily calculated. This list is by no means exhaustive. In fact, biunitary invariance is preserved under addition and multiplication, thus one can easily generate new ensembles. We do not present results for products of the ensembles considered below, solely due to the fact that the expressions for $O_2(z_1,z_2)$ become lengthy.
\begin{itemize}
\item \textit{Ginibre.} As a cross-check of correctness of our formula, let us  first consider the Ginibre ensemble. Its spectral density is uniform on the unit disk, therefore $F(r)=2\pi \int_{0}^{r}s \frac{\theta(1-s)}{\pi} ds$ is equal to 1 for $r>1$ and $F(r)=r^2$ for $r\leq 1$. Substitution to \eqref{eq:ORdiagonal} reproduces the result derived earlier \eqref{eq:Ginibre}.

\item \textit{Induced Ginibre}~\cite{IndGin}. Let us consider a rectangular $N\times M$ matrix $X$ (without loss of generality, $M>N$) with iid Gaussian entries. There exists an $M\times M$ unitary matrix $U$ so that $Y=XU$ can be represented in the block form $Y=(X',0)$. The right $N\times(M-N)$ block consists of zeros, while $X'$ is called the induced Ginibre matrix. In the limit $N,M\to\infty$ with $\alpha=\frac{M-N}{N}$ fixed, its radial cdf reads
\begin{equation}
F(r)=\left\lbrace \begin{array}{ccl}
0 & \mbox{for} & r<\sqrt{\alpha} \\
r^2-\alpha & \mbox{for} & \sqrt{\alpha}<r<\sqrt{\alpha+1} \\
1 & \mbox{for} & r>\sqrt{1+\alpha}
\end{array}\right.
\end{equation}

Substitution into \eqref{eq:ORdiagonal}  yields, after some algebra
\begin{equation}
O_{Ind}(z_1,z_2)=\frac{1}{\pi^2}\frac{(1+\alpha-z_1\zb_2)(\alpha-z_1\zb_2)}{z_1\zb_2 |z_1-z_2|^4}.
\end{equation}
The Ginibre Ensemble corresponds to $\alpha=0$.

\item \textit{Truncated Unitary}~\cite{TruncatedUnitary}. Let us consider a $(N+L)\times(N+L)$ random unitary matrix with a pdf given by the Haar measure on $U(N+L)$ and remove its last $L$ rows and columns. The radial cdf of the remaining square matrix in the limit $N,L\to\infty$, with $\kappa=\frac{L}{N}$ fixed, reads $F(r)=\kappa \frac{r^2}{1-r^2}$ for $r<(1+\kappa)^{-1/2}$ and $1$ otherwise~\cite{ProductsLargeN}. Therefore the two-point eigenvector function reads
\begin{equation}
O_{TU}(z_1,z_2)=\frac{1}{\pi^2}\frac{-1+z_1\zb_2(1+\kappa)}{|z_1-z_2|^4}.
\end{equation}

\item \textit{Spherical Ensemble} Consider the product $Y=X_1X_2^{-1}$, where $X_1$ and $X_2$ are Ginibre matrices. Its radial cdf reads $F(r)=\frac{r^2}{1+r^2}$ and its spectrum is unbounded~\cite{HaagerupSchultz}. This ensemble is beyond the assumptions made for the derivation of \eqref{eq:ORdiagonal}. Nevertheless, motivated by the successful application of these methods for the one-point correlation function in this ensemble~\cite{CondNum}, we assume the correctness of our formulas and calculate the two-point function
\begin{equation}
O_{Sph}(z_1,z_2)=\frac{1}{\pi^2}\frac{-1}{|z_1-z_2|^4}.
\end{equation}

\item \textit{Product of two Ginibre} We consider a matrix $Y=X_1X_2$, where $X_1$ and $X_2$ are Ginibre matrices. The radial cdf of $Y$ is $F(r)=\min(r,1)$, thus

\begin{equation}
O_{prod}(z_1,z_2)=-\frac{1}{\pi^2}\frac{2(|z_1|+|z_2|)(z_1\zb_2+|z_1z_2|)-|z_1+z_2|^2-4|z_1z_2|}{4|z_1z_2| |z_1-z_2|^4}. \label{eq:ProdGinibres}
\end{equation}
\end{itemize}

\subsection{Pseudohermitian matrix}

Let us consider the product $X=AB$ of two Hermitian matrices $A,B$. The product is not Hermitian, $X^{\dagger}=BA\neq X$, but if one of the matrices, let us say $A$, is positive definite, $X$ is isospectral to the Hermitian matrix $A^{1/2}BA^{1/2}$, thus $X$, despite its non-Hermiticity, has a real spectrum. The diagonalising matrix is, however, not unitary, resulting in non-orthogonality of eigenvectors. Such matrices can be toy-models for more complicated physical system described by Hamiltonians which are not Hermitian but possesses parity-time (PT) symmetry~\cite{BenderPTReview}. The most interesting models have a parameter which controls how far the system is from breaking of the symmetry. At a critical value, called the exceptional point, two real eigenvalues coalesce and move away in the imaginary direction, spontaneously breaking the PT-symmetry.

As an example we consider the matrix $X=(2+ G_1)(2+ G_2)$, where $G_i$'s are independent matrices drawn from the Gaussian Unitary Ensemble, the spectral density of which in the large $N$ limit is the Wigner semicircle, $\rho_{\textup{GUE}}(x)=\frac{1}{2\pi}\sqrt{4-x^2}$, supported on the interval $[-2,2]$. This model has an exceptional point at $x=0$.

The components of the quaternionic $R$-transform of $X$   read~\cite{multiplicationNonHermitian}
\begin{align}
R_{11}&=\frac{4(1-G_{1\ob}G_{\ob 1})(1-G_{\ob\ob})^2}{\left(1+G_{1\ob}G_{\ob 1}(G_{\ob\ob}-2)-G_{\ob\ob}+G_{11}(G_{\ob\ob}+G_{1\ob}G_{\ob 1}-1)\right)^2}, \\
R_{1\ob}&=-\frac{G_{1\ob}\left[-3-G_{1\ob}G_{\ob 1}(G_{\ob\ob}-1)+G_{\ob\ob}+G_{11}(1-G_{1\ob}G_{\ob 1}+G_{\ob\ob})\right]^2}{(G_{1\ob}G_{\ob 1}-1)\left[1+G_{1\ob}G_{\ob 1}(G_{\ob\ob}-2)-G_{\ob\ob}+G_{11}(G_{\ob\ob}-1+G_{1\ob}G_{\ob 1})\right]^2}.
\end{align}
The other two components are given by the exchange of indices $1\leftrightarrow \ob$. Inserting them into \eqref{eq:ForGreensFunction} and focusing only on the holomorpic solution ($|w|=0$), we arrive at the equation for  Green's function
\begin{equation}
\frac{4}{(1-\mathfrak{g}(z))^2}+\frac{1}{\mathfrak{g}(z)}=z. \label{eq:GreenPTmodel}
\end{equation}
We choose a branch which gives the asymptotic behavior $\mathfrak{g}(z)\sim 1/z$ for large $z$. The spectrum is supported on a single interval $[0,z_{+}]$, with $z_{+}=\frac{1}{2}(11+5\sqrt{5})$. The Green's function infinitely close to the spectrum reads 
\begin{equation}
\lim_{\epsilon\to 0} \mathfrak{g}(x\pm i\epsilon)=\frac{1+2x}{3x}-\frac{1}{6x}\left(\frac{a}{r^{1/3}}(1\pm i\sqrt{3})-r^{1/3}(1\mp i\sqrt{3})\right),
\end{equation}
where $a=1+10x+x^2$ and $r=1+15x+39x^2-z^3-6\sqrt{3}x\sqrt{x+11x^2-x^3}$. The imaginary part of Green's function yields the spectral density, calculated in~\cite{WarcholMaster}. The traced product of resolvents according to \eqref{eq:twopoint} satisfies the equation
\begin{equation}
\frac{1}{\mathfrak{h}(z_1,\zb_2)}=\frac{1}{\mathfrak{g}(z_1)\bar{\mathfrak{g}}(\zb_2)}-\frac{\left(1-\mathfrak{g}(z_1))^2(1-\bar{\mathfrak{g}}(\zb_2)\right)^2}{\left[-3+\mathfrak{g}(\zb_2)+\mathfrak{g}(z_1)+\mathfrak{g}(z_1)\mathfrak{g}(\zb_2)\right]^2}, \label{eq:PTholomorphic}
\end{equation}
where $\mathfrak{g}(z_1)$ and $\mathfrak{g}(\zb_2)$ are the solutions of \eqref{eq:GreenPTmodel} with $1/z$ asymptotic behavior.

\begin{figure}
\includegraphics[width=0.49\textwidth]{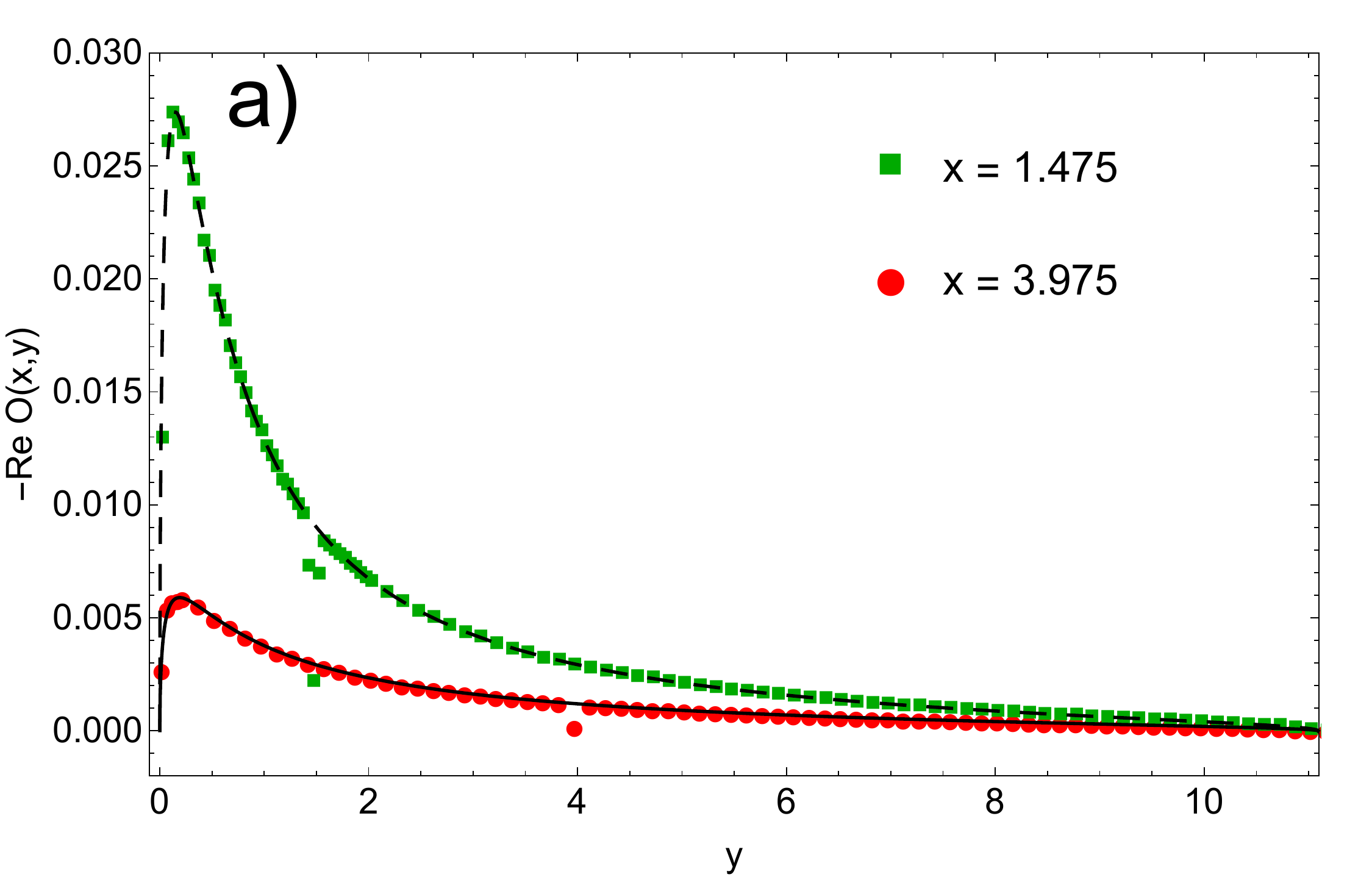} \includegraphics[width=0.49\textwidth]{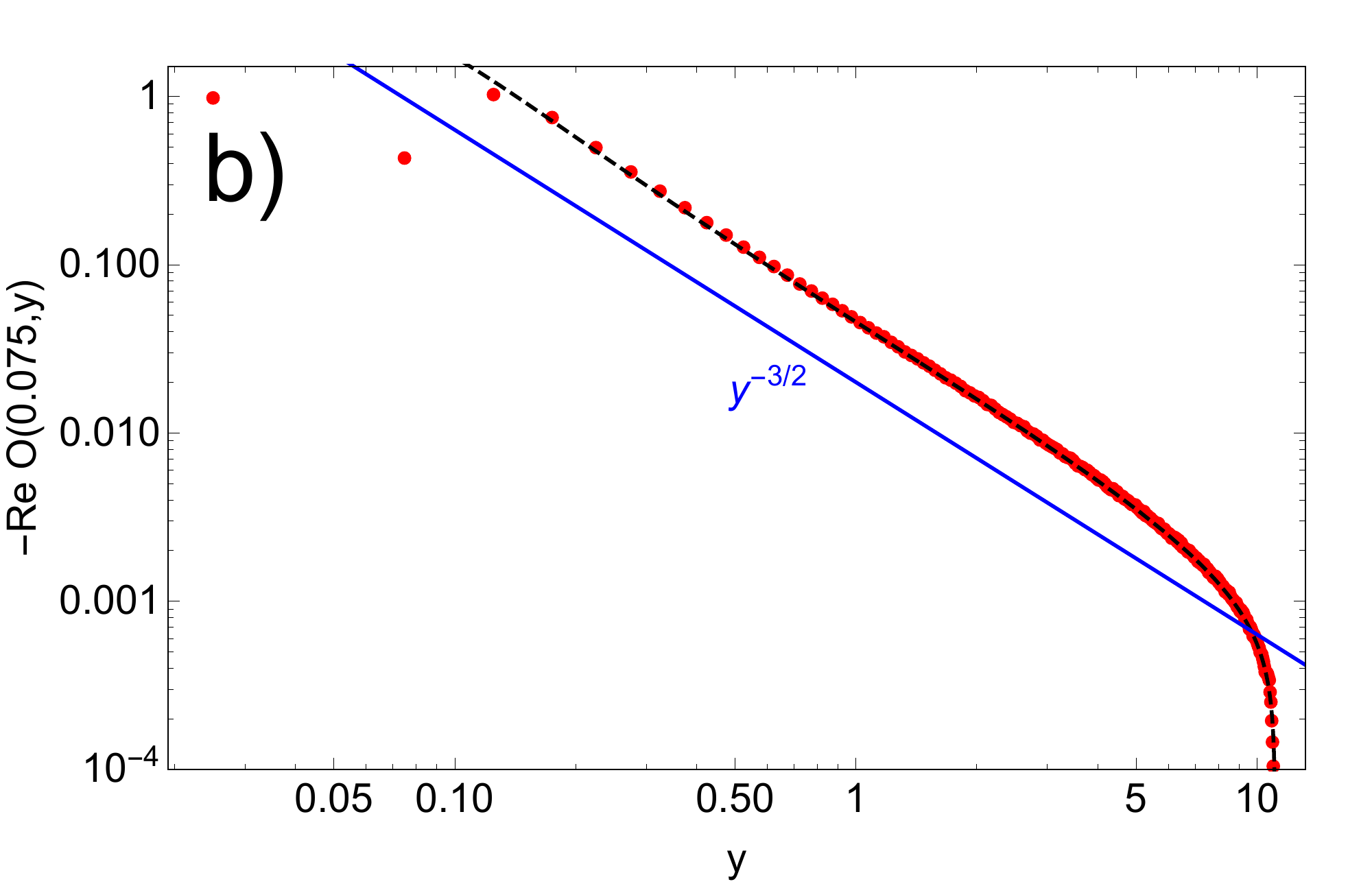}
\caption{Cross sections of the two point eigenvector correlation function $O_2(x,y)$ for a) $x=1.475$ (squares and dashed line), $x=3.975$ (circles and solid line) and b) $x=0.075$. The numerical data (points) are obtained by diagonalization of $5\cdot 10^4$ matrices of size $N=100$. Black lines are the solutions of~\eqref{eq:PTholomorphic} inserted into \eqref{eq:PTcorrelatorRule}. Interestingly, if one of the arguments is close to the exceptional point $x=0$, the large part of the function can be approximated by a power-law.
\label{Fig:PTcorrelator}
}
\end{figure}

The two-point function is calculated from \eqref{eq:CorrelationPTmodel} and its cross-sections are juxtaposed with the numerical simulations in Fig.~\ref{Fig:PTcorrelator}, showing an excellent agreement.

\section{Towards microscopic universality of eigenvectors}
\label{sec:Micro}

Random matrices show the phenomenon  of universality at certain  regions of the spectra.  In the case of Hermitian ensembles,  such universalities   appear in the bulk (the so-called  sine kernel) and  at the edges of the spectra (Airy, Bessel, Pearcey,  etc.). For a given generic Hermitian ensemble represented by $N\times N$ matrices $H$, one of the tools  for investigating the existence of universalities  are the multi-trace correlation functions
\begin{equation}
G(z_1,z_2,\ldots , z_j)=\sum_{k_1,\ldots,k_j=1}^{\infty} N^{j-2}\frac{\left< {\rm tr} H^{k_1}\ldots {\rm tr}H^{k_j}\right>_c}{z_1^{k_1+1}\ldots z_k^{k_j+1}}.
\end{equation} 
The subscript $c$ denotes the connected part. 

Such objects were studied extensively  using various techniques including loop equations~\cite{AmbjornJurkiewiczMakeenko}, Coulomb gas analogy~\cite{VivoCundenCovariance} and Feynman diagrams \cite{brezinzee1995TwoPoint,lukaszewski2008diagrammatic}. They were put into a formal mathematical formulation of the higher order freeness~\cite{SecondOrder1,SecondOrder2,SecondOrder3}. 
When the eigenvalues occupy a single interval, they obey the Ambjorn-Jurkiewicz-Makeenko universality~\cite{AmbjornJurkiewiczMakeenko}.  The divergences  of the double-trace correlation function signal the  breakdown of the $1/N$ expansion and the need to resum  the whole series and rescale its arguments. Different  universal limits are manifested as different types  of singularities. 

A natural generalization of the two-point double-trace function to the non-Hermitian setting is the connected average of two copies of the electrostatic potential~\eqref{eq:Potential} 
\begin{equation}
 F(Q,P)=\<\frac{1}{N}\ln \det (\cQ-\cX)\frac{1}{N}\ln\det (\cP-\cX)\>_c, \label{eq:logdet2}
\end{equation}
introduced in \cite{janik1997nonhermitian}, where  Gaussian models were also considered. As the quaternionic Green's function, encoding all expectations of the traces can be obtained from the potential (see ~\eqref{eq:GreenQuaternion}), the function above generates all covariances of traces
\begin{equation}
\<\frac{1}{N}\tr X^{\alpha_1}X^{\alpha_2}\ldots X^{\alpha_k} \frac{1}{N}\tr X^{\beta_1} X^{\beta_2}\ldots X^{\beta_l}\>_c, \label{eq:TwoPointDoubleTrace}
\end{equation}
being a natural extension of the second order freeness for large non-Hermitian matrices. Here $\alpha_i,\beta_j\in \{1,\dagger\}$.

As we mentioned earlier, the indices in the product of a resolvent can be contracted in two ways, see Fig.~\ref{fig:Contraction}. One of them gives access to the eigenvector correlation function, while the second one yields $F$. More precisely, $F(Q,P)= \tr L$.

Since we consider connected expectation, we may write $\ln\det(\cQ-\cX)=\ln\det(\idm-\cX\cQ^{-1})$ and use the identity $\ln\det=\tr\ln$. Then, logarithms are expanded in  power series, $\ln(1+z)=\sum_{k=1}^{\infty}\frac{z^k}{k}$, which allows for convenient calculation of Feynman diagrams. Due to the presence of traces, the baselines from $(\cX \cQ^{-1})^k$ are now drawn as two concentric rings\footnote{The rings could be equivalently drawn next to each other. This choice is just for convenience}. The dominant diagrams are the planar ones in which vertices and propagators are drawn between the two rings, but propagators connecting vertices do not encircle the inner ring (as in Fig.~\ref{fig:Contraction}), see also~\cite{brezinzee1995TwoPoint,lukaszewski2008diagrammatic}. The diagrams have an additional symmetry, namely rotating each ring leads to a new admissible diagram contributing equally. The resulting symmetry factors exactly cancel coefficients in the expansion of logarithms. 

\begin{figure}
\begin{center}
\includegraphics[width=0.49\textwidth]{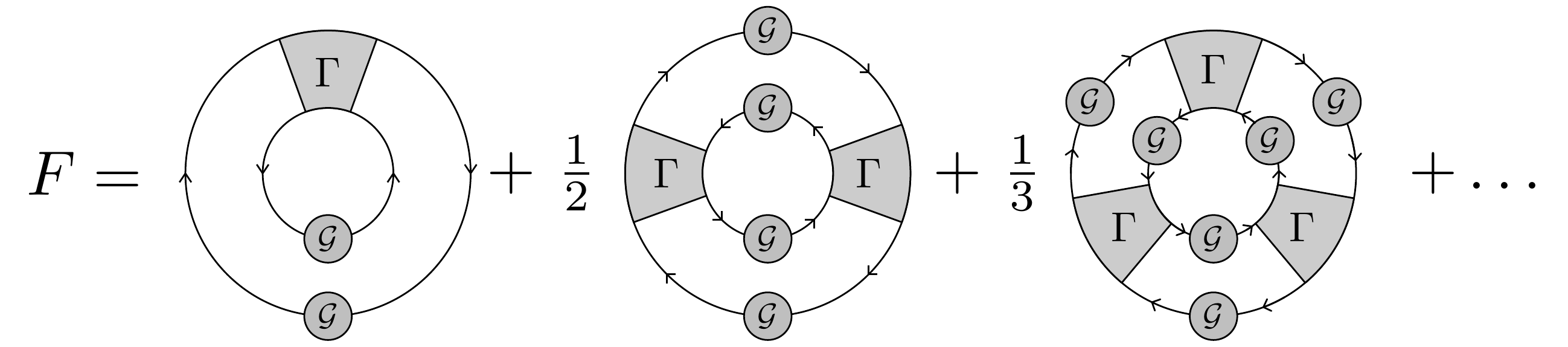}
\end{center}
\caption{Hierarchy of wheel diagrams contributing to the two-point double-trace correlation function~\eqref{eq:logdet2}, which in turn corresponds to the contraction of indices in Fig.~\ref{fig:Contraction} leading to $L$. The combinatorial factor $1/m$ corresponds to rotational symmetry and prevents overcounting the diagrams.
\label{fig:Wheels}
}
\end{figure}

Each diagram can be decomposed into $m$ segments in which $\cX$'s from two rings are connected through propagators and vertices. Segments are connected to each other through rings. As a result, each diagram looks like a wheel with $m$ spokes. It turns out that the sum of all diagrams contributing to the spoke is exactly the rung, $\Gamma$, from the ladder diagrams in Sec.~\ref{sec:Diagrams}. The $\cX$'s on ring, which are not part of a spoke can be connected with each other through propagators and vertices in any way, thus contributing to the Green's function. The general structure of such diagrams is presented in Fig.~\ref{fig:Wheels}. 

The wheel diagrams with $m$ spokes have an additional symmetry, namely they can be rotated by an angle $2\pi/m$. In order not to overcount the diagrams in the sum, we must include $1/m$ factor. Finally, we get 
\begin{equation}
N^2 F(Q,P)= \tr \sum_{m=1}^{\infty} \frac{1}{m}\left[ (G(Q)\otimes G^{T}(P))B(Q,P)\right]^{m}=-\log \det \left[\idm-(G(Q)\otimes G^T(P))B(Q,P)\right].
\end{equation}

This means that the result, derived in~\cite{janik1997nonhermitian} and used for deducing the existence of the edge universality for the \textit{spectral density}~\cite{QCDinMatter}, holds for the entire class of non-Hermitian models. 

 The two-point single-trace correlation functions encoding correlations between eigenvectors also have  their counterpart in the Hermitian case, but because of realness of the spectrum and orthogonality of eigenvectors it trivially reduces to the one-point Green's function
 \begin{equation}
 \<\frac{1}{N}\tr (z_1\idm-H)^{-1}(z_2\idm-H)^{-1}\>=\frac{\mathfrak{g}(z_1)-\mathfrak{g}(z_2)}{z_2-z_1},
 \end{equation}
thus not attracting attention. Eigenvectors of non-normal matrices are no longer trivial, making such correlation functions meaningful quantities.

In the spirit of the above analysis, one is tempted to ask if we can probe  hypothetical {\it  eigenvector universality} using similar tools. 
We would like to stress that, even in the case of the simplest Ginibre ensemble, the direct analysis of the eigenvector correlation functions is very hard. 
Whereas the finite $N$ expression for the one point function is known~\cite{mehlig2000correlator,WaltersStarr}, 
the only known non-perturbative results for the calculation of the two-point eigenvector correlation function
 are given implicitly~\cite{chalker1998correlator,mehlig2000correlator}  as
 \begin{equation}
O_2(z_1,z_2)=-\frac{N}{\pi^2 \Gamma(N)}e^{-N(|z_1|^2+|z_2|^2)}\det\left[h_{ij}\right]_{i,j=0}^{N-2}, \label{eq:Corr2Exact}
\end{equation} 
where the matrix $h$ is pentadiagonal with entries given by
\begin{equation}
h_{ij}=\frac{N^{j+3}}{\pi (j+1)!}\int d^2\lambda \bar{\lambda}^i\lambda^j\left[|z_1-\lambda|^2|z_2-\lambda|^2+\frac{1}{N}(z_1-\lambda)(\zb_2-\bar{\lambda})\right]e^{-N|\lambda|^2}.
\end{equation}

There is, however, a different possibility of inferring the existence of universality. 
Spectra of non-normal matrices  are intimately linked with the properties of their eigenvectors. The completeness relation $\sum_{k=1}^{N}\ket{R_k}\bra{L_k}=\idm$ used in the weighted density \eqref{eq:Density} leads to the sum rule $\int_\mathbb{C} d\mu(w) D(z,w)=\rho(z)$, which imposes constraints on the eigenvector correlation functions
\begin{equation}
NO_1(z)+\int\limits_{\mathbb{C}}d\mu(w)O_2(z,w)=\rho(z). \label{eq:SumRule}
\end{equation}
While the right hand side is of order 1, the one-point correlator gives a contribution of order $N$, thus there has to be a counterterm from the integral. As the region of integration is in fact compact in the large $N$ limit, the divergence can stem only from the region when $w$ is close to $z$. The exact calculations in this regime are not accessible within the diagrammatic approach, but below we give a qualitative argument that the microscopic scaling is responsible for the cancellation of divergences.

In RMT the microsopic universality can be probed on the scale of the typical distance between eigenvalues. Demanding that in the disk of radius $\delta z$ centered at $z$ we expect one eigenvalue, leads us to the scaling
\begin{equation}
w=z+\frac{u}{\sqrt{N \rho(z)}}, \label{eq:MicroScaling}
\end{equation}
where $u\sim 1$. We notice that in all examples presented in Sec.~\ref{sec:Examples} the two-point function can be expressed as 
\begin{equation}
O_2(z,w)=-\frac{1}{\pi^2}\frac{P(z,w)}{|z-w|^4}
\end{equation}
with the same behavior in the denominator. The microscoping scaling \eqref{eq:MicroScaling} inserted in the denominator produces a term $(N\rho(z))^2$, while the Jacobian of the change of variables reduces the power to one, giving the desired behavior in $N$. Moreover, by the explicit evaluation of derivatives in \eqref{eq:ORdiagonal} and the application of de l'Hospital's rule twice we get for biunitarily invariant ensembles $P(z,z)=\frac{O(z)}{\rho(z)}$, which cancels densities, eventually leaving only the one-point function, which produces the desired counterterm. We hypothesize that this phenomenon is universal across all non-Hermitian ensembles in the bulk. 

Motivated by the ubiquitousness of the $|z-w|^{-4}$ divergence in the bulk we state the conjecture that in generic non-Hermitian matrices with complex entries for all points in the bulk at which the spectral density does not develop singularities, there exists a microscopic limit
\begin{equation}
\lim_{N\to\infty}N^{-2}O_2(z+\frac{x}{\sqrt{N}},z+\frac{y}{\sqrt{N}})=O_1(z) \Phi(|x-y|),
\end{equation}
where 
\begin{equation}
\Phi(|\omega|)=-\frac{1}{\pi^2|\omega|^4}\left(1-(1+|\omega|^2)e^{-|\omega|^2}\right).\label{eq:Micro}
\end{equation}

\begin{figure}
\includegraphics[width=0.49\textwidth]{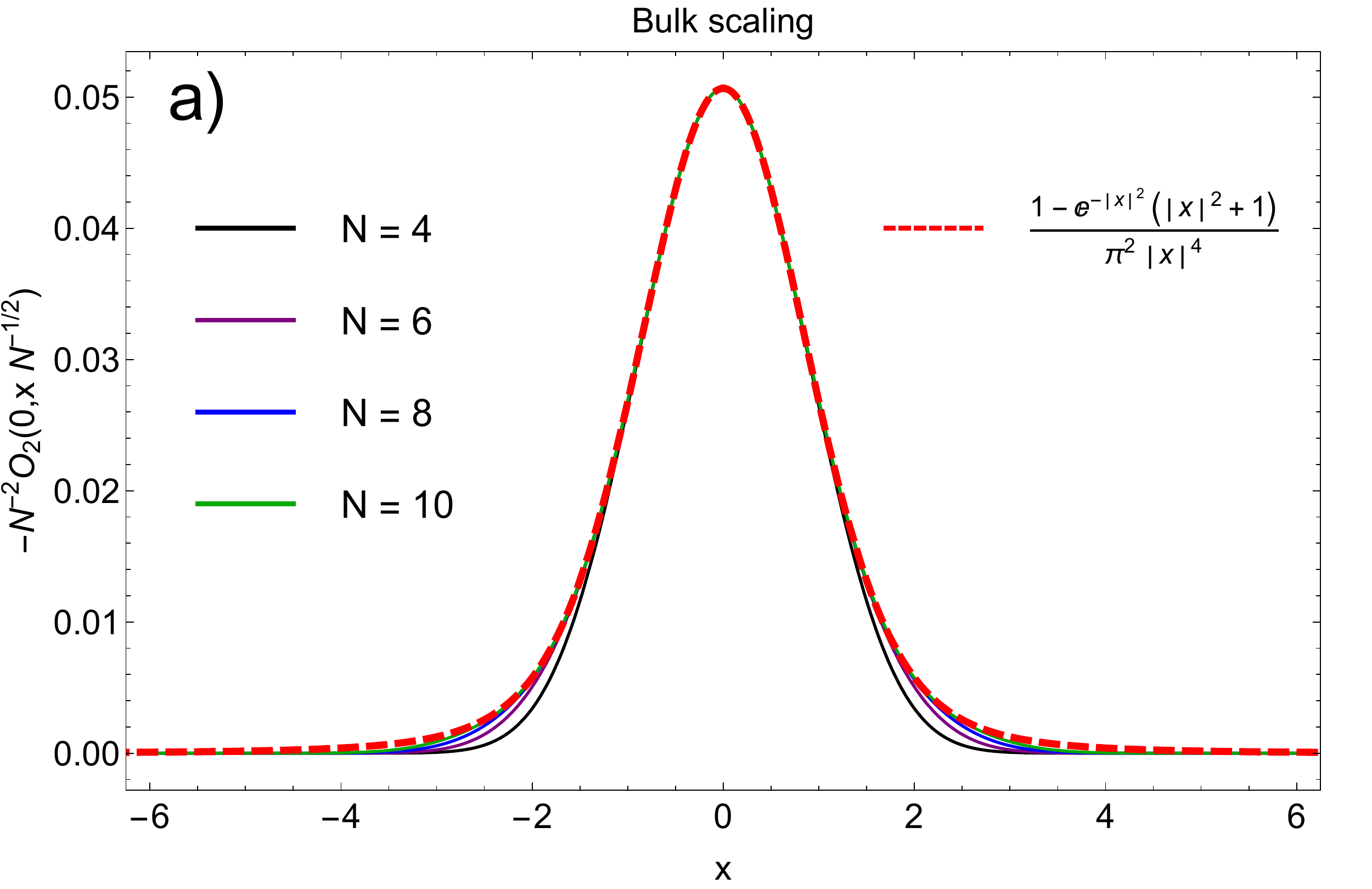}\includegraphics[width=0.49\textwidth]{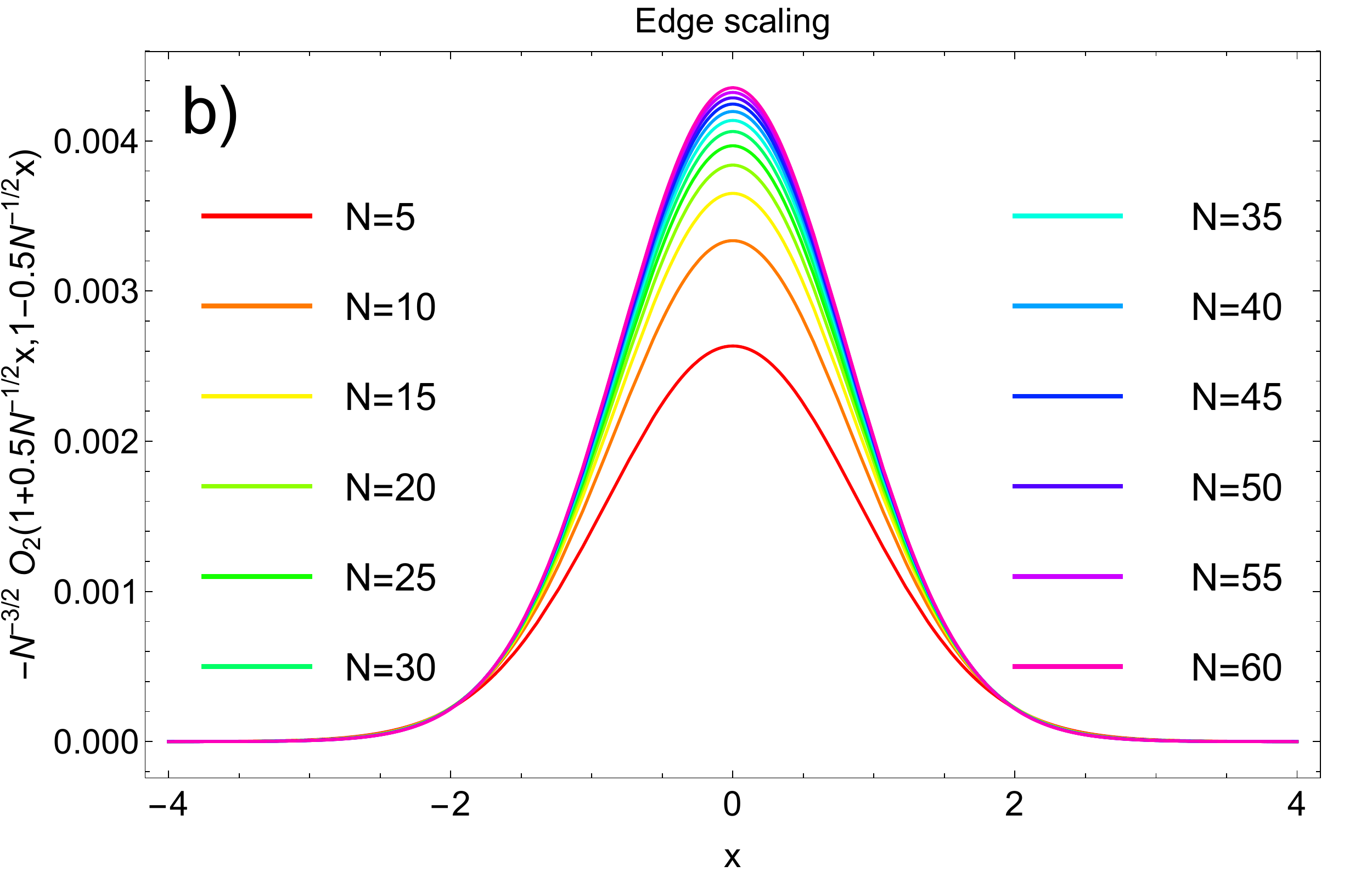}
\caption{ The two-point eigenvector correlation function $O_2(z_1,z_2)$ for the Ginibre ensemble in the microscopic a) bulk regime, $z_1=0$, $z_2=\frac{x}{\sqrt{N}}$ and b) edge regime $z_1=1+\frac{x}{2\sqrt{N}}$, $z_2=1-\frac{x}{2\sqrt{N}}$. The red dashed line is Chalker and Mehlig's exact result~\eqref{eq:Micro}. In the bulk microscopic regime $O_2$ scales as $N^2$ and at the edge as $N^{3/2}$. The rapid convergence to the limiting bulk scaling suggest that the corrections are exponentially small, while the hypothetical edge scaling seems to have $1/N$ corrections.
\label{fig:EigenvectorScaling}
}
\end{figure}

The function $\Phi$ was calculated in~\cite{chalker1998correlator,mehlig2000correlator} by evaluating $O_2(0,z)$ from the exact result~\eqref{eq:Corr2Exact} and taking the scaling limit $z=\frac{\omega}{\sqrt{N}}$. It is presented in Fig.~\ref{fig:EigenvectorScaling}a) and compared with the evaluation of the exact formula.

Interestingly, performing an analogous reasoning for the Ginibre ensemble (in which $P(z,w)=1-z\wb$) when $z$ is at the edge of the spectrum leads us to a different conclusion. When two arguments get close, the two-point function diverges, but it does so as $|z-w|^{-3}$, because $P$ also vanishes, reducing the exponent. This suggests that at the edge the two-point function scales as $N^{3/2}$ instead of $N^{2}$ in the bulk. This stays in agreement with the sum rule~\eqref{eq:SumRule}, since $O_1$ at the edge scales as $N^{1/2}$~\cite{WaltersStarr}. The limiting scaling function is not available to us, hence we evaluate numerically~\eqref{eq:Corr2Exact} and show the results in Fig.~\ref{fig:EigenvectorScaling}b). The divergence of the two-point function at the origin for the product of two Ginibre matrices~\eqref{eq:ProdGinibres} also suggests the existence of a different scaling there.
  
\section{Summary}

Using the methods of the quaternion formalism~\cite{JaroszNowakNovel} for non-Hermitian random matrices, we have proposed the explicit calculational scheme for the two-point eigenvector correlation function~\eqref{eq:CorrFuncDef}.  First, we have checked that our formalism reproduces all  known examples in the literature, i.e.  the complex Ginibre ensemble, an elliptic ensemble  and the open chaotic scattering ensemble. Second, we  considered two subclasses  of non-normal random matrices: the pseudo-hermitian  and  the biunitarily invariant ensembles, which in the large $N$ limit are described by the $R$-diagonal operators from free probability~\cite{NicaSpeicherRdiag,HiaiPetz}. In both cases we got new results for the two-point eigenvector correlation functions. In the case of the  bi-unitarily invariant ensembles, the two-point function $O_2(z,w)$  has a particularly simple form.  It is expressed solely as a function of the radial cumulative distribution function $F(r)$ and the one-point eigenvector correlation function $O_1(r)$. 

Recently, it was proven~\cite{CondNum} that  for biunitarily invariant ensmbles, $O_1(r)$ can be expressed in terms of $F(r)$ exclusively, which can be viewed as an extension of the single ring (Haagerup-Larsen) theorem~\cite{FeinbergZeeSingleRing,HaagerupLarsen}.   Combining this result with our  formalism, 
we arrive at the conclusion, that the two-point eigenvector correlation function for general biunitarily invariant ensembles in the large $N$ limit depends functionally solely  on the spectral density. Such a situation  resembles the Ambjorn-Jurkiewicz-Makeenko (known also as the Brezin-Zee) universality in the case of Hermitian random matrix models, where the two-point spectral Green's function depends solely on the one-point Green's function, irrespectively on the specific ensemble.  Mathematical formulation of such a construction is known as the second order freeness~\cite{SecondOrder1}. 
We are therefore tempted to speculate that, by combining second order freeness and freeness with amalgamation~\cite{Shlyakhtenko}, the notion of the non-orthogonality of eigenvectors can be extended into a broader context of operators in von Neumann algebras. Indeed, an  equation similar to~\eqref{eq:Master1} has recently appeared in the description of fluctuations of Gaussian block matrices~\cite{DiazMingo}.  Moreover, the diagrammatic calculations of the traced product of resolvents resemble the partition structure diagrams introduced in~\cite{haagerup2010resolvents}.

The similarity  of our result to AJM (BZ) universality  has further  consequences.  In the case of the ABJ (BZ)  universality, the singular points of the correlation functions identify the regions of the spectra where microscopic universality takes place.  This includes both the cases of the bulk  and  edge universality.  We are therefore inclined to  apply a similar  argument to our result, searching for the microscopic eigenvector  universalities.  An additional argument for the microscopic universality comes from  a constraint on eigenvector correlation function~\eqref{eq:SumRule}, as originallly noted by Walters and Starr. 
The sum rules  originating  from this  constraint strongly suggest the universal form of the microscopic two-point eigenvector correlations in analogy to a similar phenomenon for the sum rules of Dirac Euclidean operators found by Leutwyler and Smilga~\cite{LeutwylerSmilga}. The latter lead the Stony Brook  group to the discovery of the universal Bessel kernels for chiral random matrix models~\cite{ShuryakVerbaarschot,VerbaarschotZahed}. 

Our analysis, as well as explicit examples for the biunitarily invariant ensembles  calculated in Sec.~\ref{sec:ExamplesBiunitary},  point at the generic  shape of such universality, coming from the ubiquitous factor $|z-w|^4$ in the denominator. Explicitly, $O_2(z,w)= -\frac{1}{\pi^2} \frac{P(z,w)}{|z-w|^4}$, where  $\lim_{z \to w}P(z,w)=O_1(z)/\rho(z)$ yields the Petermann factor.  Such unique  behavior is responsible for the crucial cancellation of the divergent terms  in the leading order in $N$ in the sum rules. The identification of this mechanism leads us to predict the existence of the universal microscopic scaling of the eigenvector correlation function  $\Phi(|\omega|)$. Such a limit was obtained in the special case of the Ginibre ensemble~\cite{chalker1998correlator,mehlig2000correlator}. We conjecture that this universality extends to at least  biunitarily invariant random ensembles. 

Interestingly, the sum rule~\eqref{eq:SumRule}  leads also to interesting predictions at the edge. It is well known that correlations of eigenvalues of non-Hermitian matrices exhibit universal behavior at the edge, given by the error function.  Our large $N$ results for the eigenvector correlations show that  the leading singularity weakens at the edge,  $|z-w|^4 \rightarrow |z-w|^3$, leading to $N^{3/2}$ scaling of the two-point correlation function. The numerical evaluation of the implicit exact result~\eqref{eq:Corr2Exact} confirms this hypothesis, but the analytic form of the scaling function is not yet available,  even in the case of the simplest, complex Ginibre ensemble. 
 
Our results are only one step towards understanding the statistical properties of non-normal random operators and give rise to new questions. The matrix of overlaps $O_{ij}$ is the simplest invariant object. It is natural to ask what kind of non-trivial higher order invariants can be built out of eigenvectors. This problem is even more cumbersome in the light of  recent results~\cite{BourgadeDubach,FyodorovOverlap}, because the distribution of the diagonal overlap $O_{ii}$ is heavy tailed and some objects, for instance $\<O_{ii}^2\>$, do not exist. For the real Ginibre ensemble the situation is even more hopeless, since at the real axis the one-point function $O_1$ does not exist! While one expects the existence of certain correlation functions involving local averages of \textit{distinct} eigenvectors, it is unclear whether their mathematical structure simplifies as it does for spectral statistics, which form determinantal point processes. Even though  an event with two or more eigenvalues lying close to each other is unlikely to happen due to the eigenvalue repulsion, correlations between their eigenvectors do not decay, as can be seen from the microscopic scaling of $O_2$. It is therefore very appealing to study microscopic eigenvector correlations involving  more than two eigenvalues.

Although the real eigenvalues and corresponding eigenvectors of the real Ginibre ensemble are beyond the scope of perturbative techniques, we expect that the results for the two-point function remain unchanged for the eigenvectors associated with complex eigenvalues of the real Ginibre. Despite that the eigenvector overlaps are heavy-tailed, the traces of powers of $X$ and its conjugate transpose are localized around their mean value~\cite{WaltersStarr}. Such a big cancellation is possible due to the sum rule originating from the completeness relation. Based on this fact, we expect that the formula for the traced product of resolvents~\eqref{eq:twopoint} holds also for matrices with real entries.

  The issue of a hypothetical microscopic eigenvector universality for generic non-Hermitian ensembles is also of primary importance, since unraveling the unknown microscopic eigenvector correlations may give hope in the case of notorious sign problems by giving an insight into the properties of the Dirac operator in Euclidean QCD at non-zero chemical potential.   
  
\textit{Note added.} After completing this manuscript, we became aware of a recent work by Bourgade and Dubach~\cite{BourgadeDubach}, which tackles the issue of eigenvector correlations in the complex Ginibre ensemble in the bulk using different probabilistic techniques. They found the full probability of the diagonal overlap as an inverse gamma distribution and also studied the first two moments of the off-diagonal overlap. Moreover they proved that the result for the macroscopic two-point function~\eqref{eq:Ginibre} extends to mesoscopic scales.

\label{sec:Conclusions}

\section*{Acknowledgements}
WT appreciates the financial support from the Polish Ministry of Science and Higher Education through the
‘Diamond Grant’ 0225/DIA/2015/44 and the scholarship of Marian Smoluchowski Research Consortium Matter
Energy Future from KNOW funding. The authors  thank Piotr Warcho{\l}   for discussions and are grateful to Yan Fyodorov, Paul Bourgade and Guillaume Dubach for correspondence and useful remarks. The authors are indebted to Janina Krzysiak for carefully reading this manuscript and helping to bring it to the current shape.

\begin{appendices}
\section{One-point functions in Elliptic Ensemble}
\label{sec:Elliptic}
It is very instructive to show how the formalism described in Sec.~\ref{sec:Reminder} works in practice. Let us consider a non-Hermitian matrix model given by the Gaussian potential \eqref{eq:GaussianPotential}. Due to the fact that there are no vertices in this model, the only cumulants are  $c^{(2)}_{\alpha\beta}$,  given  by the propagators. This completely determines the quaternionic $R$-transform
\begin{equation}
R(Q)=\sigma^2 \left(\begin{array}{cc}
\tau Q_{11} & Q_{1\ob} \\
Q_{\ob 1} & \tau Q_{\ob\ob}
\end{array}\right).
\end{equation}
Once we perform the average over the ensemble (i.e. we know the form of $R$), we can safely remove the regularization by setting $|w|=0$ at the level of the algebraic equation~\eqref{eq:ForGreensFunction}, which in this case reads
\begin{equation}
\sigma^2\left(\begin{array}{cc}
\tau G_{11} & G_{1\ob} \\
G_{\ob 1} & \tau G_{\ob\ob}
\end{array}\right)+\frac{1}{G_{11}G_{\ob\ob}-G_{1\ob}G_{\ob 1}}\left(\begin{array}{cc}
G_{\ob\ob} & -G_{1\ob} \\
-G_{\ob 1} & G_{11}
\end{array}\right)=\left(\begin{array}{cc}
z & 0 \\
0 & \zb
\end{array}\right).
\end{equation}
Focusing on the $1\ob$ component, one gets
\begin{equation}
G_{1\ob}\left(\sigma^2-\frac{1}{G_{11}G_{\ob\ob}-G_{1\ob}G_{\ob 1}}\right)=0. \label{eq:G12Eq}
\end{equation}
There are two solutions, a trivial one $G_{1\ob}=0$ and a non-trivial one, $\sigma^2=\left(G_{11}G_{\ob\ob}-G_{1\ob}G_{\ob 1}\right)^{-1}$. Let us focus on the trivial first. Inserting $G_{1\ob}=0$ into the equation given by the $11$ component, we get $\sigma^2 \tau G_{11}+1/G_{11}=z$, with two solutions
\begin{equation}
G_{11}(z)=\frac{z\pm\sqrt{z^2-4\sigma^2\tau}}{2\sigma^2\tau}=\mathfrak{g}(z).
\end{equation}
This is the holomorphic part, valid outside the spectrum and we have to choose the branch of the solution with a minus sign for  correct asymptotic behavior at infinity $\mathfrak{g}(z)\sim 1/z$. In the holomorphic domain, the off-diagonal elements of  Green's function vanish, because the one-point eigenvector correlation function is trivially zero as there are no eigenvalues there.

Considering the non-trivial solution of \eqref{eq:G12Eq} and inserting it into the equations for $11$ and $\ob\ob$ components, we obtain a system of two linear equations
\begin{align*}
\sigma^{2}\tau G_{11}+\sigma^2G_{\ob\ob}=& z, \\
\sigma^2\tau G_{\ob\ob}+\sigma^2G_{11}= & \zb,
\end{align*}
with the solution
\begin{equation}
G_{11}(z)=\frac{\zb-z\tau}{\sigma^2 (1-\tau^2)}.
\end{equation}
The spectral density is calculated according to \eqref{eq:SpecDens}:
\begin{equation}
\rho(z,\zb)=\frac{1}{\pi}\partial_{\zb} G_{11}=\frac{1}{\pi\sigma^2(1-\tau^2)}.
\end{equation}
One can also calculate $G_{1\ob}$ and get the following formula for the one-point eigenvector correlation function from \eqref{eq:Correlator}
\begin{equation}
O_1(z)=\frac{1}{\pi \sigma^2}\left(1-\frac{|z-\zb \tau|^2}{\sigma^2(1-\tau^2)^2}\right).
\end{equation}
The boundary of the spectrum can be calculated in two ways: by requiring that the holomorphic and non-holomorphic solutions match at the boundary or by imposing vanishing of the one-point eigenvector correlation function. Both methods give
\begin{equation}
\frac{x^2}{(1+\tau)^2}+\frac{y^2}{(1-\tau)^2}=\sigma^2,
\end{equation}
which is the equation for the ellipse with semi-axes $\sigma(1+\tau)$ and $\sigma(1-\tau)$, hence the name of the ensemble.

\section{Quantum scattering ensemble}
\label{sec:QuantumScattering}
Let us see how the procedure for determining the rung of the ladder works in practice. We consider the quantum scattering ensemble~\cite{HAAKE} given by
\begin{equation}
X=H+i\gamma \Gamma,
\end{equation}
where $H$ is a $N\times N$ complex matrix with Gaussian entries of zero mean and variance $\< |H_{kl}|^2\>=N^{-1}\delta_{kl}$ and $\Gamma=\sum_{a=1}^{M} V^a(V^a)^{\dagger}$. The components of  $N$-dimensional vectors $V^a$ are complex Gaussians with variance $\<V^a_k \bar{V}^b_l\>=N^{-1}\delta_{kl}\delta_{ab}$.  The two-point eigenvector correlation function in the limit $M,N\to\infty$  with $M/N=m$ fixed (planar limit) was studied by Mehlig and Santer~\cite{MehligSanterQuantumScattering}. We show how this result can be rederived within this formalism in a simpler way.

$\Gamma$ is the complex Wishart matrix~\cite{WishartAntiwishart} multiplied by $m$. The planar cumulants of the Wishart matrix are stored in the Voiculecu's R-transform from free probability, which reads $R_{\Gamma}(z)=\frac{m}{1-z}$.
The considered matrix $X$ is non-Hermitian, therefore we need its quaternionic $R$-transform. Using the embedding of the complex $R$-transform into the quaternionic structure~\cite{JaroszNowakNovel}, we get $R_{\Gamma}(Q)=m(\idm_{2}-Q)^{-1}$. The Gaussian matrix is a particular instance of the elliptic ensemble corresponding to $\tau=1$, therefore $R_H(Q)=Q$. Further, $\Gamma$ is rescaled by a complex number $i\gamma$. The quaternionic $R$-transform of such a rescaled matrix is obtained from the relation~\cite{JaroszNowakNovel} $R_{i\gamma \Gamma}(Q)=gR_{\Gamma}(Qg)$, where $g=\diag(i\gamma,-i\gamma )$. As the $R$-transform is additive under addition of two matrices, we have $R_{X}(Q)=Q+m g (\idm-Qg)^{-1}$, which we then expand into a power series
\begin{equation}
R_X(Q)=Q+mg\sum_{k=0}^{\infty}(Qg)^k.
\end{equation}
Then we perform the procedure with acting derivatives on the quaternionic $R$-transform and substituting the argument, as described in Sec~\ref{sec:Diagrams}. After summing up the resulting series, we get
\begin{equation}
B^{\alpha\beta}_{\mu\nu}(Q,P)=\delta^{\alpha\beta}\delta_{\nu\mu}+m\left(g^{-1}-G(Q)\right)^{-1}_{\alpha\beta}\left(g^{-1}-G^{T}(P)\right)^{-1}_{\mu\nu}.
\end{equation} 
This can we written in matrix form as
\begin{equation}
B(Q,P)=\idm+m\left[g^{-1}-G(Q)\right]^{-1}\otimes \left[g^{-1}-G^{T}(P)\right]^{-1}.
\end{equation}
Inserting this into \eqref{eq:Master1}, we reproduce the results of~\cite{MehligSanterQuantumScattering}. Green's function is calculated from \eqref{eq:ForGreensFunction}.

\end{appendices}

\bibliographystyle{unsrt}
\bibliography{aaa}
%

\end{document}